\documentclass[]{elsarticle}

\usepackage{lineno,hyperref}

\usepackage{mathrsfs,amsmath}
\usepackage{amsfonts}

\usepackage{caption}
\usepackage{floatrow}
\usepackage{subfig}
\usepackage{appendix}

\usepackage{geometry}
\begin{document}

\begin{frontmatter}

\title{System and source identification from operational vehicle responses: \\ A novel modal model accounting for the track-vehicle interaction\tnoteref{mytitlenote}}
\tnotetext[mytitlenote]{Manuscript submitted to the \textit{Journal of Sound and Vibration}, in the present form, on April 1, 2016.}


\author{Giovanni De Filippis}

\author{Davide Palmieri}

\author{Leonardo Soria\corref{mycorrespondingauthor}}
\ead{leonardo.soria@poliba.it}

\author{and Luigi Mangialardi}


\address{Dipartimento di Meccanica, Matematica e Management, Politecnico di Bari, Viale Japigia 182, 70126 Bari, Italy}

\cortext[mycorrespondingauthor]{Corresponding author. Tel.: +390805962813}

\begin{abstract}
Operational Modal Analysis (OMA) is a powerful tool, widely used in the fields of structural identification and health monitoring, and certainly eligible for identifying the real in-operation behaviour of vehicle systems. Several attempts can be found in the literature, for which the usage of algorithms based on the classical OMA formulation has been strained for the identification of passenger cars and industrial trucks. The interest is mainly focused on the assessment of suspension behaviour and, thus, on the identification of the so-called vehicle rigid body modes. But issues arise when the operational identification of a vehicle system is performed, basically related to the nature of the loads induced by the roughness of rolling profiles. The forces exerted on the wheels, in fact, depending on their location, are affected by time and/or spatial correlation, and, more over, do not fit the form of white noise sequences. Thus, the nature of the excitation strongly violate the hypotheses on which the formulation of classical OMA modal model relies, leading to pronounced modelling errors and, in turn, to poorly estimated modal parameters. In this paper, we develop a specialised modal model, that we refer to as the Track-Vehicle Interaction Modal Model, able to incorporate the character of road/rail inputs acting on vehicles during operation. Since in this novel modal model the relationship between vehicle system outputs and modal parameters is given explicitly, the development of new specific curve fitting techniques, in the time-lag or frequency domain, is now possible, making available simple and cost-effective tools for vehicle operational identification. More over, a second, but not less important outcome of the proposed modal model is the usage of the resulting techniques for the indirect characterisation of rolling surface roughness, that can be used to improve comfort and safety.
\end{abstract}


\end{frontmatter}



\section{Introduction} 

The identification of the real in-operation behaviour is of utmost importance in all the stages of the vehicle design process, from model updating, extensively used as a tool for improving the accuracy of dynamics simulations, to optimisation strategies and techniques, product life-cycle management implementations, and the verification of controllers' performance \cite{Gillespie1992,Milliken1995,Guiggiani2014}. In this context, a wheeled vehicle, cruising at constant speed, on a certain road profile, is a common working occurrence for designers, in which the vehicle is mainly subjected to the external loads exerted by the interaction with the rolling rough surface. From an experimental point of view, since unknown steady-state inputs are applied to the system, Operational Modal Analysis (OMA), even referred to as in-Operation or Output-only Modal Analysis, is certainly eligible to perform the system structural identification, moving from output data only~\cite{Hermans1999,Brincker2010}. Specifically, in the case of linear systems subjected to operational loads, OMA allows for estimating the so-called modal parameters, that is the resonance frequencies, the damping ratios, the mode shapes, and the operational reference vectors, that can be utilised, in turn, to obtain a mathematical model of the system under test, generally referred to as the modal model~\cite{Heylen2007}.

Ordinary vehicle systems generally exhibit a very complex dynamics, mainly owing to nonlinearities related to shock absorbers' operation and to the kinematics imposed by the suspension design~\cite{Wallaschek1990,Surace1992}. Thus, modal models synthesised under the real operational loadings actually represent an equivalent linearisation of the tested nonlinear systems around interesting and representative working points~\cite{Caughey1963,Hagedorn1988}. Different speeds, manoeuvres, as cruising or constant radius cornering, the presence of payloads, and of different adopted designs of suspension result in different equivalent linearised models.

A novel, but even challenging application of OMA is being proposed in the framework of vehicle ride dynamics~\cite{Cossalter2004,Soria2012}. Since modal parameter estimation allows for quantifying the overall damping associated to the vehicle rigid-body modes, OMA of vehicle systems can be employed as a tool for assessing the performance of different suspension systems. More over, in the case of semi-active and active suspensions, the application of OMA is extremely interesting for the validation and optimisation of controllers' performance~\cite{Soria2012}.

During the last decade, several robust methods have been developed for estimating reliable modal parameters from output-only data~\cite{Parloo2003}. Basically, these methods rely on the Natural Excitation Technique (NExT) assumptions~\cite{James1992,Shen2003,Caicedo2004,Peeters2005}: (i) The unknown loads acting on the system have to fit the form of white noise sequences, (ii) in case of multi-point excitation, the external inputs are required to be strictly uncorrelated. The NExT assumptions are commonly satisfied in civil engineering applications, as is the case of high rise buildings and suspension bridges, where typically the structures are excited by environmental loadings due to traffic, wind and waves~\cite{Peeters2007}.

In this paper, we first focus on the issues that arise when an operational structural identification of a vehicle system is performed by using classical OMA based techniques. These issues are related to several aspects: (i) The presence of high modal density and closely-spaced modes; (ii) the high amount of damping owing to the presence of shock-absorbers; (iii) the nature of the loads induced by the rolling surface. Specifically, we show that the latter point implies a strong violation of the NExT assumptions. In fact, the trend of each excitation, transmitted to the vehicle through the wheel stiffness and damping, is quite far from the flat frequency shaping of white noise~\cite{Turkay2005,Andren2006}. In addition, depending on their location, the inputs acting on the wheels result affected by time and/or spatial correlation~\cite{Butkunas1966,Styles1976,Ammon1992,Bogsjo2008,Song2013,DeFilippis2013}. As a consequence, the application of classical OMA methodologies for post-processing the vehicle output responses may lead to pronounced modelling errors and, therefore, to poorly estimated modal parameters.

Thus, by assuming that the rolling surface is an homogeneous Gaussian random field~\cite{Dodds1973,Newland1993}, we formulate of a novel OMA modal model, referred to as the Track-Vehicle Interaction Modal Model (TVIMM), specialised and suitable for developing new identification procedures aimed at the estimation of the modal parameters of a vehicle system. More over, we show that the coefficients of the adopted empiric surface model can be even estimated as a further outcome of our OMA processing. Since direct measurements performed by using specific profilometers can result to be expensive, this second application of the model has relevant interest in road and railway health monitoring. In this field, in fact, the indirect characterisation of surface roughness can be employed as a tool for improving safety and comfort~\cite{Gonzalez2008}.

The rest of the paper is organised as follows. In Section~\ref{section2}, we obtain the representation of surface-induced forces exerted on a vehicle and discuss the issues arising when identification procedures based on the classical OMA formulation are used to process the output responses of vehicle systems. In Section~\ref{section3}, we propose a novel OMA formulation, specifically designed for the operational identification of vehicle systems, based on a specialised modal model accounting for the track-vehicle interaction. In Section~\ref{section4}, we offer a numerical demonstration of the developed formulation. Concluding remarks are summarised in Section~\ref{section5}. In Appendix~\ref{appendix}, we provide mathematical proof of some fundamental equations used in Section~\ref{section3}.


\section{System and source identification from operational vehicle responses}\label{section2}

The nature of loadings exerted on a vehicle by the rolling surface plays a fundamental role in the process of operational identification of the system. We here exploit the properties of homogeneous Gaussian random fields to derive a model of the surface-induced forces. By comparing the resulting surface inputs with those permitted by complying the NExT assumptions, we conclude that a specialised OMA formulation is needed, able to provide a correct modelling of the system and of the main excitation source, that is the surface roughness.


\subsection{Classical OMA approach} 

By processing the output responses of a $N$ degrees of freedom (dofs) system, OMA leads to the estimation of the modal parameters, that is the poles $\lambda_n$, the modal vectors ${{\boldsymbol{\psi }}_n} = {\left[ {\begin{array}{*{20}{c}}{{\psi _{1n}}}& \cdots &{{\psi_{Nn}}}\end{array}} \right]^T}\in {\mathbb{C}^{N \times 1}}$ and the operational reference vectors ${{\boldsymbol{\phi }}_n} = {\left[ {\begin{array}{*{20}{c}}{{\phi _{1n}}}& \cdots &{{\phi_{Nn}}}\end{array}} \right]^T}\in {\mathbb{C}^{N \times 1}}$ (with $n = 1, \ldots ,N$), where the symbol $(.)^T$ indicates matrix transposition. The computation of system's poles $\lambda_n$ is of fundamental interest as they contain information about the resonance frequencies $f_n=\omega_n/2\pi$ and the damping factors $\zeta_n$
\begin{equation}
{\lambda _n^{}},{\lambda_n^* }=  - {\zeta _n}{\omega _n} \pm {\rm{i}}{\omega _n}\sqrt {1 - \zeta _n^2},
\end{equation}
where $(.)^*$ indicates complex-conjugation. The application of methods of system identification to OMA has given rise to a large variety of estimation techniques~\cite{Heylen2007,Peeters2001}, comprising, generally, several steps. The most advanced procedures are based on the usage of the modal model, commonly utilised to complete the identification step and for validating the results of the estimation process~\cite{Peeters2005}. Recently, a new approach has been proposed, allowing for the computation of modal parameters directly from the modal model~\cite{El-Kafafy2015}.

In the case of classical OMA, the modal model is formulated both in the time-lag $\tau$ and in the angular frequency $\omega = 2 \pi f$ domains, by referring to correlation functions and power spectral densities (PSDs), respectively. Thus, by considering a system of external forces relying on the NExT assumptions, the input PSD matrix ${{\boldsymbol{S}}_f(\omega)}\in {\mathbb{C}^{{N} \times {N}}}$ can be written as
\begin{equation}\label{Eq:NExT}
{{\boldsymbol{S}}_f(\omega)} = \left[ {\begin{array}{*{20}{c}}
{{S_1}}&{}&0\\
{}& \ddots &{}\\
0&{}&{{S_{{N}}}}
\end{array}} \right]. 
\end{equation}
In fact, since the inputs are required to be strictly uncorrelated white noise sequences, the matrix in Eq.~\eqref{Eq:NExT} has only constant entries along the main diagonal. With referring to a $N$ dofs system subjected to the operational loadings Eq.~\eqref{Eq:NExT}, the formulation of the modal model can be given in terms of output correlation matrix $\boldsymbol{R}_q(\tau) \in {\mathbb{R}^{{N} \times {N}}}$ and output PSD matrix $\boldsymbol{S}_q(\omega) \in {\mathbb{C}^{{N} \times {N}}}$, as
\begin{equation}\label{Eq:NExTCorrMat}
{{\boldsymbol{R}}_q}\left( \tau  \right) = \sum\limits_{n = 1}^{2N} {{\boldsymbol{\phi }_n}\boldsymbol{\psi }_n^T{e^{+{\lambda _n}\tau }}} h\left( \tau  \right) + {\boldsymbol{\psi }_n}\boldsymbol{\phi }_n^T{e^{ - {\lambda _n}\tau }}h\left( { - \tau } \right),
\end{equation}
\begin{equation}\label{Eq:NExTPSDMat} 
\boldsymbol{S}_q(\omega ) = \sum\limits_{n = 1}^{2N} {\frac{{\boldsymbol{\phi}_n\boldsymbol{\psi} _n^{T}}}{{{\text{i}}\omega  - {\lambda _n}}} + \frac{{\boldsymbol{\psi}_n\boldsymbol{\phi}_n^{T}}}{{ - {\text{i}}\omega  - {\lambda _n}}}},
\end{equation}
where $h(.)$ indicates the Heaviside step function.



\subsection{Random fields for surface profiles modelling} 

Generally, in case of OMA, input loads are uncontrollable and, in addition, remain unmeasured. For this reason, the analyst has to verify that the operational loads are suitable to adequately and effectively excite the system in the frequency range of interest. 

A vehicle system in steady-state working conditions is subjected to both external and on-board excitations~\cite{Gillespie1992}. On-board forces are related to the operation of rotating parts and are specifically imputable to engine operation and to non-uniformities in assemblies and components of the driveline. These forces introduce harmonics in the spectra, whose frequencies are expected to depend on the engine rotating velocity and, in turn, on the vehicle speed. External loads are mainly related to surface roughness, that represent the only present source able to generate an effective broad band excitation. Specifically, the cut-off frequency of the input spectrum depends on roughness spatial frequency content, vehicle velocity and wheel dynamic behaviour.

Random fields for surface modelling have extensively been studied and the usage of different approaches has emerged~\cite{Dodds1973,Newland1993}. Basically, the following assumptions are commonly adopted: (i) the surface roughness is an homogeneous random field; (ii) the height of the asperities satisfies to a zero-mean Gaussian distribution; (iii) the pavement unevenness is an ergodic random process. Homogeneity implies that the statistical properties of surface roughness are independent on spatial observations. This assumption has implications similar to stationarity for one-dimensional random processes. The second assumption ensures that the output responses of a linear system subjected to pavement excitations satisfy, as well, to a zero-mean Gaussian distribution. Ergodicity guarantees that average is equal to expectation calculated over the whole ensemble. 

In simple track-vehicle interaction models, the unilateral contact point hypothesis is often utilised to describe the forces transmitted to the vehicle through the wheel stiffness and damping. This assumption is made without loss of generality, by considering that the distributed contact in the wheel-pavement interface acts as a low-pass filter, whose bandwidth is governed by the contact interface itself~\cite{Sun2006}. In this case, the surface roughness can be supposed to be a single-track random process, describing the longitudinal profile along the wheel path in the travelling direction~\cite{Turkay2005}. Generally, an empiric parametric model is adopted to fit the measured PSD, and the captured surface roughness is classified or employed in the simulations. The non-smoothed PSD is often approximated by a simple function, involving the usage of only few parameters. A literature survey on existing different approximations for longitudinal road profiles is presented in Ref.~\cite{Andren2006}.

When more than one profile is needed, the ordinary coherence function is introduced to express the relationship between multiple tracks~\cite{ISO8608}. Considering two parallel surface profiles $d_i(x)$ and $d_j(x)$, with $x$ the longitudinal spatial variable, separated by a distance $W_p$, the ordinary coherence function is defined in the angular spatial frequency domain $\nu$ as
\begin{equation}\label{Eq:coherence}
{\it{\Gamma}_{p}}\left( \nu  \right) = \frac{{\left| {{S_{{d_i}{d_j}}}\left( \nu  \right)} \right|}}{{\sqrt {{S_{{d_i}}}\left( \nu  \right){S_{{d_j}}}\left( \nu  \right)} }},
\end{equation}
where $S_{d_i}(\nu)$ and $S_{d_j}(\nu)$ are the auto-PSDs of $d_i(x)$ and $d_j(x)$, respectively, and $S_{{d_i}{d_j}}(\nu)$ is the cross-PSD between the two profiles. By definition, the coherence function is a real even function ranging from 0 to 1. As a consequence, two separated surface profiles perfectly overlap at wavelengths corresponding to amplitude values of the coherence function equal to 1. The expression of cross-PSD is obtained from Eq.~\eqref{Eq:coherence} as
\begin{equation}
{S_{{d_i}{d_j}}}\left( \nu  \right) = {\it{\Gamma}_{p}}\left( \nu  \right)\sqrt {{S_{{d_i}}}\left( \nu  \right){S_{{d_j}}}\left( \nu  \right)} {{\rm{e}}^{ - {\rm{i}}{\beta _{p}}\left( \nu  \right)}},
\end{equation}
where $\beta_{p}(\nu)$ is the difference between the phases related to $d_i(x)$ and $d_j(x)$.

Based on the homogeneity assumption, the following properties can be exploited for simplifying the description of multiple tracks~\cite{Dodds1973,Newland1993}: \\ (i) The auto-PSDs related to parallel surface profiles are coinciding and, thus, equal to the same function 
\begin{equation}\label{Eq:hom1}
{S_{{d_i}}}\left( \nu  \right) = {S_{{d_j}}}\left( \nu  \right) = {S_d}\left( \nu  \right);
\end{equation}
(ii) the cross-PSD between two parallel surface profiles is a real and even function depending on auto-PSD and coherence function 
\begin{equation}\label{Eq:hom2}
{S_{{d_i}{d_j}}}\left( \nu  \right) = {\it{\Gamma}_{p}}\left( \nu  \right){S_d}\left( \nu  \right),
\end{equation}
%
%
%
in which it is specifically useful to notice the cancellation of the phase difference $\beta_{p}(\nu)$ in Eq.~\eqref{Eq:hom2}. 


\subsection{Problem statement}

\begin{figure}[htbp]
\centering
\includegraphics[width=9cm]{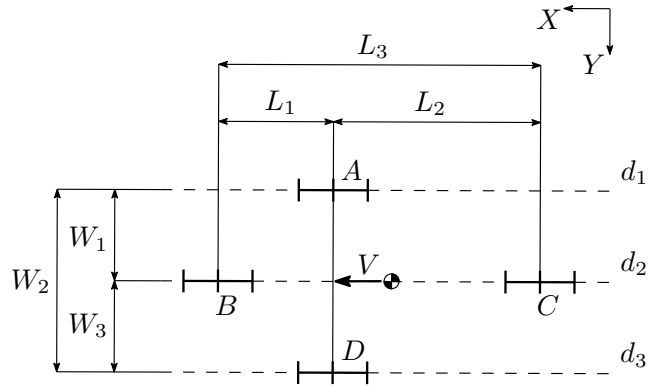}
\caption{Geometry of a generic $N_t$-wheel vehicle system.}\label{Fig:vehicle_4}
\end{figure}

We here consider the generic $N_t$-wheel vehicle system represented in Fig.~\ref{Fig:vehicle_4}, with $N_t = 4$, even if the following considerations can be easily extended to systems equipped with $N_t \neq 4$ wheels. The considered geometry includes $N_w = 3$ different trackwidths $W_p$ (with $p=1,\ldots,N_w$) and $N_l = 3$ different wheelbases $L_l$ (with $l=1,\ldots,N_l$). We hypothesise that the vehicle travels at constant velocity $V$ on an homogeneous Gaussian surface. By writing Eqs. from~\eqref{Eq:coherence} to~\eqref{Eq:hom2} in the angular frequency domain $\omega = \nu V$, we obtain the following PSD matrix of surface-induced displacements $\boldsymbol{S}_{r}(\omega) \in {\mathbb{C}^{{N_t} \times {N_t}}}$ 
\begin{align}\label{Eq:DispPSDMat}
{{\boldsymbol{S}}_{r}}\left( \omega  \right) & = \left[ {\begin{array}{*{20}{c}}
{{S_{{d_A}}}\left( \omega  \right)}&{{S_{{d_A}{d_B}}}\left( \omega  \right)}& \cdots &{{S_{{d_A}{d_D}}}\left( \omega  \right)}\\
{S_{{d_A}{d_B}}^*\left( \omega  \right)}&{{S_{{d_B}}}\left( \omega  \right)}&{}& \vdots \\
 \vdots &{}& \ddots & \vdots \\
{S_{{d_A}{d_D}}^*\left( \omega  \right)}& \cdots & \cdots &{{S_{{d_D}}}\left( \omega  \right)}
\end{array}} \right] \nonumber \\
& = \left[ {\begin{array}{*{20}{c}}
1&{{\it{\Gamma} _1}\left( \omega  \right){{\rm{e}}^{ + {\rm{i}}\omega {\tau _1}}}}&{{\it{\Gamma} _1}\left( \omega  \right){{\rm{e}}^{ - {\rm{i}}\omega {\tau _2}}}}&{{\it{\Gamma} _2}\left( \omega  \right)}\\
{{\it{\Gamma} _1}\left( \omega  \right){{\rm{e}}^{ - {\rm{i}}\omega {\tau _1}}}}&1&{{{\rm{e}}^{ - {\rm{i}}\omega {\tau _3}}}}&{{\it{\Gamma} _3}\left( \omega  \right){{\rm{e}}^{ - {\rm{i}}\omega {\tau _1}}}}\\
 \vdots &{}&1&{{\it{\Gamma} _3}\left( \omega  \right){{\rm{e}}^{{\rm{ + i}}\omega {\tau _2}}}}\\
{{\it{\Gamma} _2}\left( \omega  \right)}& \cdots & \cdots &1
\end{array}} \right]{S_d}\left( \omega  \right),
\end{align}
where $\tau_l  = L_l /V$ indicate the time-delays between the inputs acting on different axles. 

To describe the forces transmitted to the vehicle through the wheel stiffness and damping, we first introduce the following static gain matrix $\boldsymbol{G}_{fr} \in {\mathbb{R}^{{N} \times {2 N_t}}}$ 
\begin{equation}\label{Eq:StaticGainMat}
{{\boldsymbol{G}}_{fr}} = \left[ {\begin{array}{*{20}{c}}
{{\textbf{\textit{0}}^{\left( {N - {N_t}} \right) \times {N_t}}}}&{{\textbf{\textit{0}}^{\left( {N - {N_t}} \right) \times {N_t}}}}\\
{{{\boldsymbol{K}}_t}}&{{{\boldsymbol{C}}_t}}
\end{array}} \right],
\end{equation}
where $\boldsymbol{K}_t$ and $\boldsymbol{C}_t \in {\mathbb{R}^{{N_t} \times {N_t}}}$ are diagonal matrices in which stiffness and damping terms are, respectively, collected. We, second, account for the PSDs of the derived processes, in addition to Eq.~\eqref{Eq:DispPSDMat}, in case tyre damping is included in the wheel model. We, finally, obtain the following representation for the input PSD matrix related to surface-induced forces ${{\boldsymbol{S}}_f(\omega)}\in {\mathbb{C}^{{N} \times {N}}}$
\begin{equation}\label{Eq:SiFPSDMat}
{{\boldsymbol{S}}_f}\left( \omega  \right) = {\boldsymbol{G}_{fr}}\left( {{\boldsymbol{{D}}}\left( \omega  \right) \otimes {{\boldsymbol{S}}_r}\left( \omega  \right)} \right)\boldsymbol{G}_{fr}^T 
\quad \mbox{with} \quad 
{\boldsymbol{D}}\left( \omega  \right) = \left[ {\begin{array}{*{20}{c}}
1&{{\rm{i}}\omega }\\
{ - {\rm{i}}\omega }&{{\omega ^2}}
\end{array}} \right],
\end{equation}
where the matrix $\boldsymbol{D}(\omega) \in {\mathbb{C}^{{4} \times {4}}}$ allows for taking into account the contributions related to road-induced velocities and the symbol $\otimes$ denotes the Kronecker product. We stress that the sign of the imaginary part of Eq.~\eqref{Eq:SiFPSDMat} depends on the adopted definition of correlation function, where a change of definition in the time-lag domain leads to a complex-conjugate expression in the frequency domain.

By comparing Eqs.~\eqref{Eq:NExT} and~\eqref{Eq:SiFPSDMat}, we conclude that for a vehicle system subjected to operational loadings, the modal model resulting from the classical OMA approach (Eqs.~\eqref{Eq:NExTCorrMat} and~\eqref{Eq:NExTPSDMat}) is no more valid. The arisen issues are mainly related to the nature of the loads induced by the rolling surface, that violates the NExT assumptions. First, we notice that forces applied to each wheel are basically coloured excitations, with magnitude inversely proportional to the frequency raised to a certain power, and shaping depending on the auto-PSD of the road profile ($S_d(\omega)$ Eq.~\eqref{Eq:hom1}) and on the wheel parameters ($\boldsymbol{G}_{fr}$ Eq.~\eqref{Eq:StaticGainMat}). In addition, being systems equipped with more than one wheel, forces are affected by time and/or spatial correlation: (i) wheels mounted on the same axle are subjected to spatially correlated inputs ($\it{\Gamma}_{p}(\omega)$ in Eq.~\eqref{Eq:DispPSDMat}); (ii) wheels travelling on the same path and located on different axles are subjected to time correlated inputs (${{\rm{e}}^{ \pm {\rm{i}}\omega {\tau _l}}}$ in Eq.~\eqref{Eq:DispPSDMat}); (iii) wheels travelling on separated paths and located on different axles are subjected to time and spatially correlated inputs ($\it{\Gamma}_{p}\left( \omega  \right) {{\rm{e}}^{ \pm {\rm{i}}\omega {\tau _l}}}$ in Eq.~\eqref{Eq:DispPSDMat}).

Owing to the aforementioned effects, the application of classical OMA methodologies to vehicle responses may lead to pronounced modelling errors and, therefore, to poorly estimated modal parameters. The formulation of a specialised modal model for correctly describing the track-vehicle interaction is required and, in turn, a specific OMA formulation is needed to understand how Eqs.~\eqref{Eq:NExTCorrMat} and~\eqref{Eq:NExTPSDMat} are modified. We stress that a modal model providing the relationship between the generic output of the system and its modal parameters in an explicit form is needed for the formulation of a whatever procedure for modal parameter estimation based on suitable curve fitting algorithms. We more over comment that since operational vehicle responses incorporate information about the surface roughness, the identification procedures based on this novel OMA formulation would allow for estimating, in addition to modal parameters, even the coefficients of the adopted empiric surface models.


\section{The Track-Vehicle Interaction Modal Model}\label{section3}

We here utilise a 7 dofs system ($N=7$), generally referred to as the full-car model, to introduce the theoretical background of the proposed Track-Vehicle Interaction Modal Model. To obtain the analytical expression of the generic system output, we solve a Duhamel integral in modal coordinates.  Moving from Eq.~\eqref{Eq:SiFPSDMat}, we compute the correlation matrix of surface-induced forces and achieve the TVIMM formulation by using properties of convolution integrals and Fourier transform.


\subsection{Full-car model}

\begin{figure}[htbp]
\centering
\includegraphics[width=13cm]{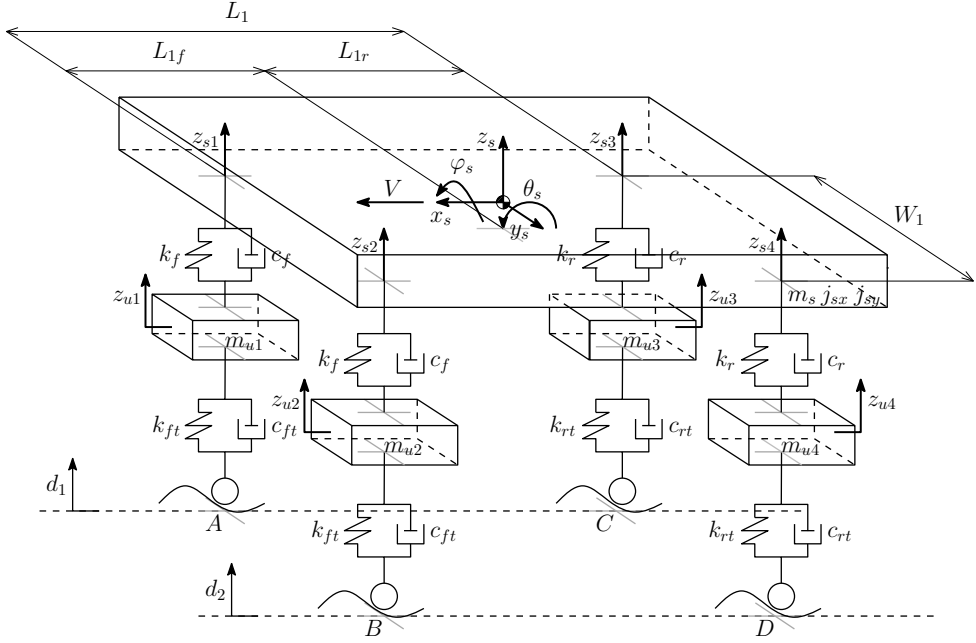}
\caption{Schematic representation of the full-car model.}\label{Fig:Full-car}
\end{figure}

The full-car model (Fig.~\ref{Fig:Full-car}) is the simplest mathematical description of a four-wheel vehicle ($N_t=4$), whose predicted output responses incorporate all the effects of the issues discussed in Section~\ref{section2}, which make no longer possible the usage of classical OMA methodologies for post-processing. Since full-car model offers a good trade-off between model complexity and accuracy, this linear lumped-parameters system is commonly utilised for simulating the ride dynamics of passenger and race cars~\cite{Milliken1995,Guiggiani2014}. The model preserves the multi-input nature of road excitation and, different from half-car model, allows for evaluating the contributions to vehicle responses due to roll disturbance produced by two parallel tracks. The geometry comprises one single trackwidth $W_1$ and one single wheelbase $L_1$. In Fig.~\ref{Fig:Full-car}, we indicate the dofs of the model by using the following notation: $z_s$, $\varphi_s$, and $\theta_s$ represent the heave, the roll and the pitch rigid body motions of the sprung mass (the body); $z_{s1}$, $z_{s2}$, $z_{s3}$, $z_{s4}$ denote the vertical displacements at the four corners of the sprung mass (strut mounts to body); $z_{u1}$, $z_{u1}$, $z_{u3}$, $z_{u4}$ are the rattle displacements of the unsprung masses (the wheels). More over, with regards to system parameters, $m_s$, $j_{sx}$, $j_{sy}$ represent the mass and the moments of inertia associated with the body; $m_{u1}$, $m_{u2}$, $m_{u3}$, $m_{u4}$ denote the unsprung masses; $k_f$, $k_r$, $c_f$, $c_r$ are the stiffness and damping coefficients of the absorbers; $k_{ft}$, $k_{rt}$, $c_{ft}$, $c_{rt}$ indicate the stiffness and damping coefficients of the tyres;
$L_{1f}$, $L_{1r}$ are the distances of the front and rear axle, respectively, from the center of gravity of the unsprung mass, the sum of which equals the wheelbase.

  

\subsection{Duhamel integral in modal coordinates}

Since vehicles are generally non-proportional damping systems, we adopt a representation in state-space form. Thus, we recast the motion equations of a $N$ dofs system into an equivalent set of 2$N$ first-order differential equations. Specifically, by denoting with $\boldsymbol{q}(t) \in {\mathbb{R}^{N \times 1}}$ the vector of Lagrangian coordinates and with $\boldsymbol{f}(t) \in {\mathbb{R}^{N \times 1}}$ the vector of external loads, we write the well-known set of system dynamics equations as
\begin{equation}\label{Eq:TD_Model}
\boldsymbol{M} \: \ddot{\boldsymbol{q}}(t) + \boldsymbol{C} \: {\dot{\boldsymbol{q}}}(t) + \boldsymbol{K} \: \boldsymbol{q}(t) = \boldsymbol{f}(t),
\end{equation}
where $t$ is the time variable, and $\boldsymbol{M}$, $\boldsymbol{C}$, and $\boldsymbol{K} \in {\mathbb{R}^{N \times N}}$ are the mass, damping, and stiffness matrices, respectively. By adding the following further set of differential equations
\begin{equation}
{\boldsymbol{M}{\dot{\boldsymbol{q}}}(t)} - {\boldsymbol{M}{\dot{\boldsymbol{q}}}(t)} = {\mbox{\textbf{\textit{0}}}}^{N \times N},
\end{equation}
and by making the substitutions
\begin{equation}
{\boldsymbol{{x}}(t)} = \left[ {\begin{array}{*{20}{c}}
{\boldsymbol{q}(t)}\\
{{\dot{\boldsymbol{q}}}(t)}
\end{array}} \right]
\qquad
{{\boldsymbol{\dot{x}}}}(t) = \left[ {\begin{array}{*{20}{c}}
{{\dot{\boldsymbol{q}}}(t)}\\
{{\ddot{\boldsymbol{q}}}(t)}
\end{array}} \right]
\qquad
{\boldsymbol{{u}}(t)}  = \left[ {\begin{array}{*{20}{c}}
{\boldsymbol{f}(t)}\\
 {\mbox{\textbf{\textit{0}}}}^{N \times 1}
\end{array}} \right],
\end{equation}
we obtain the set of system dynamics equations in state-space form 
\begin{equation}\label{Eq:SS_Form}
{\boldsymbol{P}} {{\boldsymbol{\dot{{x}}}}}(t)  + {\boldsymbol{Q}}{\boldsymbol{{x}}(t)} = {\boldsymbol{u}(t)},
\end{equation}
where $\boldsymbol{P}$ and $\boldsymbol{Q} \in {\mathbb{R}^{2N \times 2N}}$ are partitioned as 
\begin{equation}\label{Eq:SSMatrices}
{\boldsymbol{P}} = \left[ {\begin{array}{*{20}{c}}
{\boldsymbol{C}}&{\boldsymbol{M}}\\
{\boldsymbol{M}}&{\textbf{\textit{0}}^{N \times N}}
\end{array}} \right]
\qquad
{\boldsymbol{Q}} = \left[ {\begin{array}{*{20}{c}}
{\boldsymbol{K}}&{\textbf{\textit{0}}^{N \times N}}\\
{\textbf{\textit{0}}^{N \times N}}&{ - {\boldsymbol{M}}}
\end{array}} \right].
\end{equation}
Here, the product $-{\boldsymbol{P}^{ - 1}}{{\boldsymbol{Q}}}$ leads to the so-called state matrix $\boldsymbol{A}$. The eigenvalue decomposition of the state matrix $\boldsymbol{A}$ is
\begin{equation}\label{Eq:StateMatrix}
{{\boldsymbol{A}}} = {\boldsymbol{V}\boldsymbol{\it{\Lambda}}}{{\boldsymbol{V}}^{ - 1}},
\end{equation}
where ${\boldsymbol{\it{\Lambda}}} \in \mathbb{C}^{\: 2N \times 2N}$ is the eigenvalue matrix, containing the system poles 
\begin{equation}
{{\boldsymbol{\it{\Lambda}}}} = {\rm{diag}}\left( {{\lambda_1}, \ldots ,{\lambda _N}},{{\lambda_1}^*, \ldots ,{\lambda _N}^*} \right),
\end{equation}
with diag$(.)$ denoting a diagonal matrix, while ${\boldsymbol{V}} \in \mathbb{C}^{\: 2N \times 2N}$ is the eigenvector matrix, having the following structure
\begin{equation}\label{Eq:ModalVectors}
\boldsymbol{V} = \left[ {\begin{array}{*{20}{c}}
{{{\boldsymbol{\psi }}_1}}& \cdots &{{{\boldsymbol{\psi }}_N}}&{{\boldsymbol{\psi }}_1^*}& \cdots &{{\boldsymbol{\psi }}_N^*}\\
{{\lambda _1}{{\boldsymbol{\psi }}_1}}& \cdots &{{\lambda _N}{{\boldsymbol{\psi }}_N}}&{\lambda _1^*{\boldsymbol{\psi }}_1^*}& \cdots &{\lambda _N^*{\boldsymbol{\psi }}_N^*}
\end{array}} \right].
\end{equation} 
We decouple the equations of motion by using the following coordinate transformation from the physical to the modal space
\begin{equation}\label{Eq:Mod_Coords}
{\boldsymbol{x}(t)} = \boldsymbol{V} {\boldsymbol{p}(t)} 
\quad \Rightarrow \quad
\left[ {\begin{array}{*{20}{c}}
{\boldsymbol{q}(t)}\\
{{\dot{\boldsymbol{q}}}(t)}
\end{array}} \right] = \sum\limits_{n = 1}^{2N} {\left[ {\begin{array}{*{20}{c}}
{{{\boldsymbol{\psi }}_n}}\\
{{\lambda _n}{{\boldsymbol{\psi }}_n}}
\end{array}} \right]} {p_n}\left( t \right),
\end{equation}
where ${{\boldsymbol{p}}}(t) = {\left[ {\begin{array}{*{20}{c}}{{p _1}(t)}& \cdots &{{p_{2N}}(t)}\end{array}} \right]^T}\in {\mathbb{R}^{N \times 1}}$ is the modal state vector. By substituting Eq.~\eqref{Eq:Mod_Coords} in Eq.~\eqref{Eq:SS_Form} and pre-multiplying both sides by $\boldsymbol{V}^T$, we obtain 2$N$ independent differential equations, that we collect in the following compact form
\begin{equation}\label{Eq:Ma_Mb}
{{\boldsymbol{M}}_{a}}{\dot{\boldsymbol{p}}(t)} + {{\boldsymbol{M}}_{b}}{\boldsymbol{p}(t)} = {\boldsymbol{V}^T}{\boldsymbol{u}}(t),
\end{equation}
where $\boldsymbol{M}_a$ and $\boldsymbol{M}_b \in {\mathbb{C}^{2N \times 2N}}$ are two diagonal matrices generally referred to as modal a and modal b. Specifically,
\begin{equation}
{{\boldsymbol{M}}_{a}} = {\boldsymbol{V}^T}{\boldsymbol{P}}\boldsymbol{V} = {\text{diag}}\left( {{m_{a1}}, \ldots ,{m_{aN}},m_{a1}^*, \ldots ,m_{aN}^*} \right)
\end{equation}
and
\begin{equation}
{{\boldsymbol{M}}_{b}} = {\boldsymbol{V}^T}{\boldsymbol{Q}}\boldsymbol{V} = {\text{diag}}\left( {{m_{b1}}, \ldots ,{m_{bN}},m_{b1}^*, \ldots ,m_{bN}^*} \right),
\end{equation}
while
\begin{equation}
{{\boldsymbol{V}}^T}{\boldsymbol{u}}\left( t \right) = {\left[ {\begin{array}{*{20}{c}}
{{{\boldsymbol{\psi }}_1}}& \cdots &{{{\boldsymbol{\psi }}_N}}&{{\boldsymbol{\psi }}_1^*}& \cdots &{{\boldsymbol{\psi }}_N^*}
\end{array}} \right]^T}{\boldsymbol{f}}\left( t \right).
\end{equation}
By considering the generic system motion equation of set Eq.~\eqref{Eq:Ma_Mb} referred to the $n$-th mode of vibration,
\begin{equation}
{m_{an}}{{\dot p}_n}\left( t \right) + {m_{bn}}{p_n}\left( t \right) = {{\boldsymbol{\psi }}_n}^T{\boldsymbol{f}}(t) \quad \mbox{with} \quad {m_{bn}} =  - {\lambda _n}{m_{an}},
\end{equation}
and by assuming zero initial conditions, we retrieve the solution in the form 
\begin{equation}\label{Eq:Duhamel_Integral}
{p_n}\left( t \right) = \frac{{{{\boldsymbol{\psi }}_n}^T}}{{{m_{an}}}}\int\limits_{ - \infty }^t {{\boldsymbol{f}(\epsilon)}{{\rm{e}}^{{\lambda _n}\left( {t - \varepsilon } \right)}}{\rm{d}}\varepsilon}.
\end{equation}
Eq.~\eqref{Eq:Duhamel_Integral}, usually written in physical coordinates, is the well-known Duhamel integral. By combining Eqs.~\eqref{Eq:Mod_Coords} and~\eqref{Eq:Duhamel_Integral}, we obtain the system response in terms of the physical Lagrangian coordinates
\begin{equation}\label{Eq:Generic_Output}
{\boldsymbol{q}}(t) = \sum\limits_{n = 1}^{2N} {{{\boldsymbol{\psi }}_n}{p_n}\left( t \right)}  = \sum\limits_{n = 1}^{2N} {\frac{{{{\boldsymbol{\psi }}_n}{{\boldsymbol{\psi }}_n}^T}}{{{m_{an}}}}\int\limits_{ - \infty }^t {{\boldsymbol{f}}(\rho){{\rm{e}}^{{\lambda _n}\left( {t - \rho } \right)}}{\rm{d}}\rho } }.
\end{equation}
We comment that Eq.~\eqref{Eq:Duhamel_Integral} can be interpreted as a convolution integral encompassing the impulse response matrix $\boldsymbol{g}(t) \in {\mathbb{R}^{N \times N}}$ of the system under study
\begin{equation}
{\boldsymbol{g}}\left( t \right) = \sum\limits_{n = 1}^{2N} {\frac{{{{\boldsymbol{\psi }}_n}{{\boldsymbol{\psi }}_n}^T}}{{{m_{an}}}}{{\rm{e}}^{{\lambda _n}t}}}.
\end{equation}



\subsection{Input and output correlation functions}

By considering the generic outputs $q_i(t)$ and $q_j(t)$ in Eq.~\eqref{Eq:Generic_Output}, evaluated at the separated time instants $t$ and $t+\tau$, respectively,
\begin{equation}
{q_i}\left( t \right) = \sum\limits_{n = 1}^{2N} {\frac{{{{\psi _{in}}{\boldsymbol{\psi }}_n}^T}}{{{m_{an}}}}\int\limits_{ - \infty }^t {{\boldsymbol{f}}\left( \rho  \right){{\rm{e}}^{{\lambda _n}\left( {t - \rho } \right)}}{\rm{d}}\rho } },
\end{equation}
and
\begin{equation}
{q_j}\left( {t + \tau} \right) = \sum\limits_{m = 1}^{2N} {\frac{{{{\psi _{jm}}{\boldsymbol{\psi }}_m}^T}}{{{m_{am}}}}\int\limits_{ - \infty }^{t + \tau} {{\boldsymbol{f}}\left( \sigma  \right){{\rm{e}}^{{\lambda _m}\left( {t + \tau - \sigma } \right)}}{\rm{d}}\sigma } },
\end{equation}
where the following relations hold between the dummy integration variables
\begin{equation}
\rho  \to t    \quad  \mbox{and} \quad \sigma  \to t + \tau \quad \Leftrightarrow \quad  \sigma  = \rho  + \tau  \quad  \mbox{and} \quad  \tau = \sigma  - \rho,
\end{equation}
we derive the resulting output cross-correlation function 
\begin{align}\label{Eq:R_qi_qj_1}
{R_{{q_i}{q_j}}}\left( \tau \right) & = {\rm{E}}\left[ {{q_i}\left( t \right){q_j}\left( {t + \tau} \right)} \right] \nonumber \\
& = \sum\limits_{n = 1}^{2N} {\sum\limits_{m = 1}^{2N} {\frac{{{\psi _{in}}{\psi _{jm}}}}{{{m_{an}}{m_{am}}}}\int\limits_{ - \infty }^{t + \tau} {\int\limits_{ - \infty }^t {{{\boldsymbol{\psi }}_n}^T{\rm{E}}\left[ {{\boldsymbol{f}}\left( \rho  \right){{\boldsymbol{f}}^T}\left( {\rho  + \tau} \right)} \right]{{\boldsymbol{\psi }}_m}{{\rm{e}}^{{\lambda _n}\left( {t - \rho } \right)}}{{\rm{e}}^{{\lambda _m}\left( {t + \tau - \sigma } \right)}}{\rm{d}}\rho {\rm{d}}\sigma } } } } \nonumber \\
& = \sum\limits_{n = 1}^{2N} {\sum\limits_{m = 1}^{2N} {\frac{{{\psi _{in}}{\psi _{jm}}}}{{{m_{an}}{m_{am}}}}\int\limits_{ - \infty }^{t + \tau} {\int\limits_{ - \infty }^t {{{\boldsymbol{\psi }}_n}^T{{\boldsymbol{R}}_{f}}\left( \tau \right){{\boldsymbol{\psi }}_m}{{\rm{e}}^{{\lambda _n}\left( {t - \rho } \right)}}{{\rm{e}}^{{\lambda _m}\left( {t + \tau - \sigma } \right)}}{\rm{d}}\rho {\rm{d}}\sigma } } } },
\end{align}
where the symbol $\rm{E}[.]$ indicates the expectation computed over the ensemble and ${\boldsymbol{R}}_{f}(\tau) \in {\mathbb{R}^{N \times N}}$ is the input correlation matrix related to the external forces, which, by definition, is the inverse Fourier transform of the input PSD matrix Eq.~\eqref{Eq:SiFPSDMat}. 

By particularising to the case of full-car model (Fig.~\ref{Fig:Full-car}) the stiffness and damping entries of the static gain matrix $\boldsymbol{G}_{fr}$ Eq.~\eqref{Eq:StaticGainMat}
\begin{equation}
\boldsymbol{K}_t = {\rm{diag}}(k_{ft},k_{ft},k_{rt},k_{rt}) 
\qquad
\boldsymbol{C}_t = {\rm{diag}}(c_{ft},c_{ft},c_{rt},c_{rt}), 
\end{equation}
we decompose the PSD matrix of surface-induced displacements $\boldsymbol{S}_{r}(\omega)$ through the following summation  
\begin{equation}\label{Eq:DispPSDMat_2}
{{\boldsymbol{S}}_r}\left( \omega  \right) = {S_d}\left( \omega  \right)\sum\limits_{p = 0}^{{N_w}} {{{\it \Gamma }_p}\left( \omega  \right)} {{\boldsymbol{\it{\Delta}}}_p}\left( \omega  \right) \quad \mbox{with} \quad {\it \Gamma }_0(\omega)=1,
\end{equation}
where $N_w = 1$ in the case of full car model and the matrices ${\boldsymbol{\it{\Delta}}}_{0}(\omega)$ and ${\boldsymbol{\it{\Delta}}}_{1}(\omega) \in {\mathbb{C}^{N_t \times N_t}}$ are defined as 
\begin{equation}
{{\boldsymbol{\it{\Delta}}}_{0}}(\omega) = \left[ {\begin{array}{*{20}{c}}
1&0&{{{\rm{e}}^{ - {\rm{i}}\omega {\tau _1}}}}&0\\
0&1&0&{{{\rm{e}}^{ - {\rm{i}}\omega {\tau _1}}}}\\
{{{\rm{e}}^{{\rm{ + i}}\omega {\tau _1}}}}&0&1&0\\
0&{{{\rm{e}}^{{\rm{ + i}}\omega {\tau _1}}}}&0&1
\end{array}} \right],
\end{equation}
\begin{equation}
{{\boldsymbol{\it{\Delta}}}_{1}}(\omega) = \left[ {\begin{array}{*{20}{c}}
0&1&0&{{{\rm{e}}^{ - {\rm{i}}\omega {\tau _1}}}}\\
1&0&{{{\rm{e}}^{ - {\rm{i}}\omega {\tau _1}}}}&0\\
0&{{{\rm{e}}^{{\rm{ + i}}\omega {\tau _1}}}}&0&1\\
{{{\rm{e}}^{{\rm{ + i}}\omega {\tau _1}}}}&0&1&0
\end{array}} \right].
\end{equation}
By substituting Eq.~\eqref{Eq:DispPSDMat_2} in Eq.~\eqref{Eq:SiFPSDMat}, we obtain the Fourier pair ${\boldsymbol{S}}_{f}(\omega)$ and ${\boldsymbol{R}}_{f}(\tau)$ in the case of interest 
\begin{equation}\label{Eq:SiFPSDMat_2}
{{\boldsymbol{S}}_f}\left( \omega  \right) = {S_d}\left( \omega  \right)\sum\limits_{p = 0}^{{N_w}} {{\it {\Gamma}_p}\left( \omega  \right){{\boldsymbol{G}}_{fr}}\left( {{\boldsymbol{D}}\left( \omega  \right) \otimes {{\boldsymbol{\it{\Delta}}}_p}\left( \omega  \right)} \right){\boldsymbol{G}}_{fr}^T} \quad \mbox{with} \quad {\it \Gamma }_0(\omega)=1,
\end{equation}
and
\begin{equation}\label{Eq:SiFCorrMat}
{{\boldsymbol{R}}_f}\left( \tau  \right) = {R_d}\left( \tau  \right) * \sum\limits_{p = 0}^{{N_w}} {{{H}_p}\left( \tau  \right) * \left( {{{\boldsymbol{G}}_{fr}}\left( {{\boldsymbol{d}}\left( \tau  \right) \otimes {{\boldsymbol{\delta }}_p}\left( \tau  \right)} \right){\boldsymbol{G}}_{fr}^T} \right)} \quad \mbox{with} \quad {\it H }_0(\tau)=1,
\end{equation}
where $R_d(\tau)$ is the auto-correlation function associated to surface profiles and $H_p(\tau)$ (with $p \ne 0$) represents the inverse Fourier transform (IFT) of the ordinary coherence function $\it{\Gamma_p}(\omega)$ referred to two parallel tracks. The terms $\boldsymbol{d}(\tau) \in {\mathbb{R}^{{4} \times {4}}}$, ${\boldsymbol{\delta }}_{0}(\tau)$ and ${\boldsymbol{\delta }}_{1}(\tau) \in {\mathbb{R}^{{N_t} \times {N_t}}}$ correspond to a representation in the time-lag domain of the matrices included in Eq.~\eqref{Eq:SiFPSDMat_2}, that is
\begin{equation}   
{\boldsymbol{d}(\tau)} = \left[ {\begin{array}{*{20}{c}}
1&{\frac{{{\rm{d}}\left( . \right)}}{{{\rm{d}}\tau}}}\\
{ - \frac{{{\rm{d}}\left( . \right)}}{{{\rm{d}}\tau}}}&{ - \frac{{{{\rm{d}}^2}\left( . \right)}}{{{\rm{d}}{\tau^2}}}}
\end{array}} \right],
\end{equation}
\begin{equation}\label{Eq:DeltaMat1}
{{\boldsymbol{\delta }}_{0}}(\tau)  = \left[ {\begin{array}{*{20}{c}}
{\delta \left( \tau \right)}&0&{\delta \left( {{\tau-\tau_1}} \right)}&0\\
0&{\delta \left( \tau \right)}&0&{\delta \left( {{\tau-\tau_1}} \right)}\\
{\delta \left( {{\tau+\tau_1}} \right)}&0&{\delta \left( \tau \right)}&0\\
0&{\delta \left( {{\tau+\tau_1}} \right)}&0&{\delta \left( \tau \right)}
\end{array}} \right],
\end{equation}
\begin{equation}\label{Eq:DeltaMat2}
{{\boldsymbol{\delta }}_{1}}(\tau)  = \left[ {\begin{array}{*{20}{c}}
0&{\delta \left( \tau \right)}&0&{\delta \left( {{\tau-\tau_1}} \right)}\\
{\delta \left( \tau \right)}&0&{\delta \left( {{\tau-\tau_1}} \right)}&0\\
0&{\delta \left( {{\tau+\tau_1}} \right)}&0&{\delta \left( \tau \right)}\\
{\delta \left( {{\tau+\tau_1}} \right)}&0&{\delta \left( \tau \right)}&0
\end{array}} \right],
\end{equation}
where $\delta(.)$ indicates the Dirac delta function, and the symbols $\frac{{{\text{d}}(.)}}{{{\text{d}}\tau }}$ and $\frac{{{{\text{d}}^2}(.)}}{{{{\text{d}}^2}\tau }}$ denote the first and second order derivative operators, respectively. By combining Eqs.~\eqref{Eq:SiFCorrMat} and~\eqref{Eq:R_qi_qj_1}, we achieve the final expression of the cross-correlation function ${R_{{q_i}{q_j}}}\left( \tau \right)$
\begin{equation}\label{Eq:R_qi_qj_2}
{R_{{q_i}{q_j}}}\left( \tau  \right) = {R_d}\left( \tau  \right) * \sum\limits_{p = 0}^{{N_w}} {{{H}_p}\left( \tau  \right) * }{\sum\limits_{n = 1}^{2N} {\sum\limits_{m = 1}^{2N} {\frac{{{\psi _{in}}{\psi _{jm}}}}{{{m_{an}}{m_{am}}}}\int\limits_{ - \infty }^{t + \tau } {\int\limits_{ - \infty }^t {{\boldsymbol{\psi }}_n^T{{\boldsymbol{G}}_{fr}}\left( {{\boldsymbol{d}}\left( \tau  \right) \otimes {{\boldsymbol{\delta }}_p}\left( \tau  \right)} \right){\boldsymbol{G}}_{fr}^T{{\boldsymbol{\psi }}_m}{{\rm{e}}^{{\lambda _n}\left( {t - \rho } \right)}}{{\rm{e}}^{{\lambda _m}\left( {t + \tau  - \sigma } \right)}}{\rm{d}}\rho {\rm{d}}\sigma } } } } }.
\end{equation}


\subsection{Modal model in the time-lag and frequency domains}

Moving from Eq.~\eqref{Eq:R_qi_qj_2}, by means of calculations detailed in Appendix~\ref{appendix}, we obtain the relationship between the operational modal parameters ($\lambda_n$, $\psi_{in}$, $\alpha^p_{in}$, $\beta^{pl}_{in}$, $\chi^{pl}_{in}$) and the generic output cross-correlation function ${R_{{q_i}{q_j}}}(\tau)$. We here rephrase the expression of the output correlation matrix ${\boldsymbol{R}}_q\left( \tau  \right)$, in the following compact matrix notation
\begin{equation}\label{Eq:CorrMat}
{{\boldsymbol{R}}_q}\left( \tau  \right) = \sum\limits_{n = 1}^{2N} {{\boldsymbol{\bar \varphi}_n}(\tau)\boldsymbol{\psi }_n^T{e^{+{\lambda _n}\tau }}} + {\boldsymbol{\psi }_n}\boldsymbol{\bar \varphi}_n^T(-\tau){e^{ - {\lambda _n}\tau }},
\end{equation}
where the terms $\boldsymbol{\bar \varphi}_n(\tau)$ are the lag-dependent operational reference vectors, defined as   
\begin{equation}\label{Eq:CorrMat2}
{{\boldsymbol{\bar \varphi }}_n}\left( \tau  \right) = {R_d}\left( \tau  \right) * {\sum\limits_{p = 0}^{{N_W}} {{H _p}\left( \tau  \right) * \left( {\boldsymbol{\alpha }_n^p h\left( \tau  \right) + \sum\limits_{l = 1}^{{N_L}} {{\boldsymbol{\beta }_n^{pl}{{\rm{e}}^{ + {\lambda _n}{\tau _l}}}h\left( {\tau  + {\tau _l}} \right) + {\boldsymbol{\chi }}_n^{pl}{{\rm{e}}^{ - {\lambda _n}{\tau _l}}}h\left( {\tau  - {\tau _l}} \right)}} } \right)} } \quad \mbox{with} \quad H_0(\tau)=1.
\end{equation}
By Fourier transforming Eq.~\eqref{Eq:CorrMat}, we obtain the expression of the output PSD matrix ${\boldsymbol{S}}_q\left(\omega\right)$
\begin{equation}\label{Eq:PSDMat} 
\boldsymbol{S}_q(\omega ) = \sum\limits_{n = 1}^{2N} {\frac{{\boldsymbol{\varphi}_n(\omega)\boldsymbol{\psi} _n^{T}}}{{{\text{i}}\omega  - {\lambda _n}}} + \frac{{\boldsymbol{\psi}_n\boldsymbol{\varphi}_n^{T}(-\omega)}}{{ - {\text{i}}\omega  - {\lambda _n}}}},
\end{equation}
where the terms $\boldsymbol{\varphi}_n(\omega)$ are the frequency-dependent operational reference vectors, defined as   
\begin{equation}\label{Eq:PSDMat2}
{\boldsymbol{\varphi} _n}\left( \omega  \right) = {S_d}\left( \omega  \right) {\sum\limits_{p = 0}^{{N_W}} {{{\mathit \Gamma}_p}\left( \omega  \right)\left( {\boldsymbol{\alpha }_n^p + \sum\limits_{l = 1}^{{N_L}} {\boldsymbol{\beta }_n^{pl}{{\rm{e}}^{{\rm{ + i}}\omega {\tau _l}}} + {\boldsymbol{\chi }}_n^{pl}{{\rm{e}}^{ - {\rm{i}}\omega {\tau _l}}}}} \right)} } \quad \mbox{with} \quad {\mathit \Gamma}_0(\omega)=1.
\end{equation}

The modal model that we introduce through Eqs.~\eqref{Eq:CorrMat} and~\eqref{Eq:PSDMat}, namely the TVIMM, incorporates a combination of several contributions: (i) $R_d(\tau)$ and $S_d(\omega)$ describe the statistical properties of the single surface profiles; (ii) ${{e^{ \pm {\lambda _n}{\tau _l}}}} h\left( {\tau \pm {\tau_l}} \right)$ and ${{\text{e}}^{ \pm {\text{i}}\omega {\tau_l}}}$ account for the time correlation between the inputs applied to wheels travelling on the same path and mounted on different axles; (iii) $H_p(\tau)$ and ${\mathit \Gamma_p}(\omega)$ encompass the spatial correlation due to loads acting on wheels travelling on separated paths. Thus, the vehicle output responses contain information on the dynamics of the system and on its interaction with the surface profiles. Not considering this fundamental last contribution from the resulting modal model (Eqs.~\eqref{Eq:CorrMat} and~\eqref{Eq:PSDMat}) increase the modelling errors of the identification procedure. As a consequence, the unknown parameters result poorly estimated.

We consider useful to stress that although Eqs.~\eqref{Eq:CorrMat} and~\eqref{Eq:PSDMat} have been obtained by only considering the rigid-body modes of the full-car model, this novel OMA formulation is of general validity and can be utilised for real-world vehicle systems. Specifically, it is able to include and describe, without loss of generality, even the presence of deformable-body modes of vibration.


\subsection{Comments on the usage of half spectra}

Several robust OMA algorithms in the frequency domain rely on the estimation of the so called half spectra~\cite{Peeters2007}, defined as 
\begin{equation}
S_{{q_i}{q_j}}^{\left(  +  \right)}\left( \omega  \right) = \sum\limits_{n = 1}^{2N} {\frac{{{\phi _{in}}{\psi _{jn}}}}{{i\omega  - {\lambda _n}}}}.
\end{equation}

The main advantages of this representation consist in: (i) The usage of lower model orders without affecting the quality of the fitting; (ii) the adoption of well-known windowing functions for reducing the effect of leakage and the influence of samples at the higher time-lags, which are the most affected by noise. 

In the case of the classical OMA formulation, half spectra $S_{q_iq_j}^{(+)}(\omega)$ are computed by Fourier transforming the correlation functions corresponding to positive time-lags
\begin{equation}
R_{{q_i}{q_j}}^{\left(  +  \right)}\left( \tau  \right) = \sum\limits_{n = 1}^{2N} {{\phi _{in}}{\psi _{jn}}{e^{ + {\lambda _n}\tau }}} h\left( \tau  \right).
\end{equation}

We here prove that in presence of time correlation this property does no longer hold. In fact, the lag-dependent operational reference vectors $\boldsymbol{\bar \varphi}_n^T(-\tau)$ do not vanish at the positive time-lags
\begin{equation}
{\bf{\bar \varphi }}_n^T\left( { - \tau } \right) = {R_d}\left( \tau  \right)*\sum\limits_{p = 0}^{{N_W}} {{H_p}\left( \tau  \right)*\sum\limits_{l = 1}^{{N_L}} {{\bf{\beta }}{{_n^{pl}}^T}{{\rm{e}}^{ - {\lambda _n}{\tau _l}}}h\left( { - \tau  + {\tau _l}} \right)} } \quad \mbox{for} \quad \tau>0,
\end{equation}
implying that the Fourier transform of Eq.~\ref{Eq:CorrMat} for $\tau > 0$ does not lead to the first fraction summation in Eq.~\ref{Eq:PSDMat}.


\section{Numerical demonstration}\label{section4}

In this section, we offer a numerical demonstration of the effectiveness of Eqs.~\eqref{Eq:CorrMat} and~\eqref{Eq:PSDMat}. Specifically, we first compute the responses of the full-car model to surface-induced excitations by using the following implicit analytical expression holding in the frequency domain
\begin{equation}\label{Eq:In-Out_MIMO}
\boldsymbol{S}_q\left( \omega  \right) = \boldsymbol{G}_{qf}^*\left( \omega  \right)\boldsymbol{S}_f\left( \omega  \right)\boldsymbol{G}_{qf}^T\left( \omega  \right),
\end{equation}
representing the well-known input-output formula~\cite{Newland1993}. Since the full-car model utilised in this study is a linear mutliple-input mutliple-output system, the output PSD matrix $\boldsymbol{S}_q\left( \omega  \right)$ can be calculated at each frequency through Eq.~\eqref{Eq:In-Out_MIMO}. We derive the frequency response function matrix $\boldsymbol{G}_{qf}(\omega) \in {\mathbb{C}^{N \times N}}$, referred to the lumped-parameters system of Fig.~\ref{Fig:Full-car}, by means of the following relation
\begin{equation}\label{Eq:FRF_Mat}
\boldsymbol{G}_{qf}\left( \omega  \right) = {\left( { - {\omega ^2}\boldsymbol{M} + {\text{i}}\omega \boldsymbol{C} + \boldsymbol{K}} \right)^{ - 1}}.
\end{equation}
Then, we compare the obtained curves with those resulting from the TVIMM. We stress that the comparison, here provided for demonstration purposes, is made between quantities computed by following two different routes: The former is based on Eq.~\eqref{Eq:In-Out_MIMO}, in which the relationship between the generic output of the system and its modal parameters in not explicitly given; the latter is based on the proposed OMA for vehicles' modal model Eqs.~\eqref{Eq:CorrMat} and~\eqref{Eq:PSDMat}, in which the relationship is instead provided in an explicit form. Only this second formulation allows for developing specialised curve fitting techniques, needed, in turn, for modal parameter estimation. We more over comment that, based on the different approximations proposed in the literature, semi-analytical expressions of $S_d(\omega)$ and $\it{\Gamma}_p(\omega)$ have to be provided, in order to form the input PSD matrix of surface-induced forces $\boldsymbol{S}_f(\omega)$ (Eq.~\eqref{Eq:SiFPSDMat}).


\subsection{Matrix formulation of the simulated vehicle model}

For the full-car model, two alternative sets of degrees of freedom can be used to describe the dynamics of the sprung mass: (i) The heave $z_s$, the roll $\varphi_s$, and the pitch $\theta_s$ or (ii) the vertical translations at the right-front $z_{s1}$, left-front $z_{s2}$, and right-rear $z_{s3}$ corners. In what follows, we analyse both the two cases by introducing two different Lagrangian coordinate vectors, respectively
\begin{equation}
{\boldsymbol{q}_1}(t) = {\left[ {\begin{array}{*{20}{c}}
{{z_s}}&{{\varphi _s}}&{{\theta _s}}&{{z_{{u_1}}}}&{{z_{{u_2}}}}&{{z_{{u_3}}}}&{{z_{{u_4}}}}
\end{array}} \right]^T}
\qquad
{\boldsymbol{q}_2}(t) = {\left[ {\begin{array}{*{20}{c}}
{{z_{s1}}}&{{z_{s2}}}&{{z_{s3}}}&{{z_{{u_1}}}}&{{z_{{u_2}}}}&{{z_{{u_3}}}}&{{z_{{u_4}}}}
\end{array}} \right]^T}.
\end{equation}
The vector of surface-induced forces $\boldsymbol{f}(t)$ is 
\begin{equation}
{\boldsymbol{f}}\left( t \right) = {\left[ {\begin{array}{*{20}{c}}
0&0&0&{{c_{ft}}{{\dot d}_A}\left( t \right) + {k_{ft}}{d_A}\left( t \right)}&{{c_{ft}}{{\dot d}_B}\left( t \right) + {k_{ft}}{d_B}\left( t \right)}&{{c_{rt}}{{\dot d}_C}\left( t \right) + {k_{rt}}{d_C}\left( t \right)}&{{c_{rt}}{{\dot d}_D}\left( t \right) + {k_{rt}}{d_D}\left( t \right)}
\end{array}} \right]^T}.
\end{equation}
The mass, damping and stiffness matrices of the considered vehicle model, related to the first set of Lagrangian coordinates ${\boldsymbol{q}_1}(t)$, are given by
\begin{equation}
{\boldsymbol{M}_1} = \text{diag}\left( {{m_s},{j_{sx}},{j_{sy}},{m_{u1}},{m_{u2}},{m_{u3}},{m_{u4}}} \right),
\end{equation}
\begin{equation}
{{\boldsymbol{C}}_1} = \left[ {\begin{array}{*{20}{c}}
{2\left( {{c_f} + {c_r}} \right)}&0&{2\left( {{c_r}{L_{1r}} - {c_f}{L_{1f}}} \right)}&{ - {c_f}}&{ - {c_f}}&{ - {c_r}}&{ - {c_r}}\\
0&{0.5\left( {{c_f} + {c_r}} \right)W_1^2}&0&{0.5\,{c_f}{W_1}}&{ - 0.5\,{c_f}{W_1}}&{0.5\,{c_r}{W_1}}&{ - 0.5\,{c_r}{W_1}}\\
{2\left( {{c_r}{L_{1r}} - {c_f}{L_{1f}}} \right)}&0&{2\left( {{c_f}L_{1f}^2 + {c_r}L_{1r}^2} \right)}&{{c_f}{L_{1f}}}&{{c_f}{L_{1f}}}&{ - {c_r}{L_{1r}}}&{ - {c_r}{L_{1r}}}\\
{ - {c_f}}&{0.5\,{c_f}{W_1}}&{{c_f}{L_{1f}}}&{{c_f} + {c_{ft}}}&0&0&0\\
{ - {c_f}}&{ - 0.5\,{c_f}{W_1}}&{{c_f}{L_{1f}}}&0&{{c_f} + {c_{ft}}}&0&0\\
{ - {c_r}}&{0.5\,{c_r}{W_1}}&{ - {c_r}{L_{1r}}}&0&0&{{c_r} + {c_{rt}}}&0\\
{ - {c_r}}&{ - 0.5\,{c_r}{W_1}}&{ - {c_r}{L_{1r}}}&0&0&0&{{c_r} + {c_{rt}}}
\end{array}} \right],
\end{equation}
\begin{equation}
{{\boldsymbol{K}}_1} = \left[ {\begin{array}{*{20}{c}}
{2\left( {{k_f} + {k_r}} \right)}&0&{2\left( {{k_r}{L_{1r}} - {k_f}{L_{1f}}} \right)}&{ - {k_f}}&{ - {k_f}}&{ - {k_r}}&{ - {k_r}}\\
0&{0.5\left( {{k_f} + {k_r}} \right)W_1^2}&0&{0.5\,{k_f}{W_1}}&{ - 0.5\,{k_f}{W_1}}&{0.5\,{k_r}{W_1}}&{ - 0.5\,{k_r}{W_1}}\\
{2\left( {{k_r}{L_{1r}} - {k_f}{L_{1f}}} \right)}&0&{2\left( {{k_f}L_{1f}^2 + {k_r}L_{1r}^2} \right)}&{{k_f}{L_{1f}}}&{{k_f}{L_{1f}}}&{ - {k_r}{L_{1r}}}&{ - {k_r}{L_{1r}}}\\
{ - {k_f}}&{0.5\,{k_f}{W_1}}&{{k_f}{L_{1f}}}&{{k_f} + {k_{ft}}}&0&0&0\\
{ - {k_f}}&{ - 0.5\,{k_f}{W_1}}&{{k_f}{L_{1f}}}&0&{{k_f} + {k_{ft}}}&0&0\\
{ - {k_r}}&{0.5\,{k_r}{W_1}}&{ - {k_r}{L_{1r}}}&0&0&{{k_r} + {k_{rt}}}&0\\
{ - {k_r}}&{ - 0.5\,{k_r}{W_1}}&{ - {k_r}{L_{1r}}}&0&0&0&{{k_r} + {k_{rt}}}
\end{array}} \right],
\end{equation}
while those related to the second set ${\boldsymbol{q}_2}(t)$, can be expressed by
\begin{equation}
{{\boldsymbol{M}}_2} = {{\boldsymbol{M}}_1}{{\boldsymbol{T}}^{ - 1}}
\qquad
{{\boldsymbol{C}}_2} = {{\boldsymbol{C}}_1}{{\boldsymbol{T}}^{ - 1}}
\qquad
{{\boldsymbol{K}}_2} = {{\boldsymbol{K}}_1}{{\boldsymbol{T}}^{ - 1}},
\end{equation}
where $\boldsymbol{T} \in {\mathbb{R}^{N \times N}}$ is the transformation matrix
\begin{equation}
{\boldsymbol{T}} = \left[ {\begin{array}{*{20}{c}}
{\left[ {\begin{array}{*{20}{c}}
1&{ - 0.5\,{W_1}}&{ - {L_{1f}}}\\
1&{0.5\,{W_1}}&{ - {L_{1f}}}\\
1&{ - 0.5\,{W_1}}&{{L_{1r}}}
\end{array}} \right]}&{{\mbox{\textbf{\textit{0}}}^{4 \times 4}}}\\
{{\mbox{\textbf{\textit{0}}}^{4 \times 4}}}&{{\mbox{\textbf{\textit{I}}}^{4 \times 4}}}
\end{array}} \right].
\end{equation}
The parameters of the full-car model employed in the simulation \cite{Guiggiani2014} are listed in Tab.~\ref{Tab:full-car_pars}. 
\begin{table}[htbp]
\centering
\begin{tabular}{|c|c|c|c|c|c|}
\hline
Quantity  & Unit & Value & Quantity & Unit & Value \\
\hline
\hline
 $W_1$ & m & 1.490 & $c_r$ & kN s/m & 1.310 \\
\hline 
$L_{1f}$ & m & 1.064 & $m_{u1}$ & kg &  57.500 \\
\hline
$L_{1r}$ & m & 1.596 &  $m_{u2}$ & kg &  57.500 \\
\hline
$m_s$  & kg & 1150 & $m_{u3}$ & kg &  57.500 \\
\hline
$j_{sx}$ & kg m\textsuperscript{2} & 530 & $m_{u4}$ & kg &  57.500 \\
\hline 
$j_{sy}$ & kg m\textsuperscript{2} & 1630 & $k_{ft}$ & kN/m & 140 \\
\hline 
$k_f$ & kN/m & 15.750 & $k_{rt}$ & kN/m & 140 \\
\hline
$k_r$ & kN/m  & 14 & $c_{ft}$ & kN s/m & 0.150 \\
\hline 
$c_f$ & kN s/m & 1.475 & $c_{rt}$ & kN s/m & 0.150 \\
\hline
\end{tabular}
\caption{Parameters of the full-car model \cite{Guiggiani2014}.}
\label{Tab:full-car_pars}
\end{table}


\subsection{Single-profile and coherence-based models for parallel road paths}

We obtain the input PSD matrix of surface-induced forces by imposing that the auto-PSD of single surface profiles $S_d(\omega)$ be equal to the ISO 8608 class C approximation. According to ISO 8608~\cite{ISO8608}, in fact, roughness level is classified from A to H, and the form of the fitted one-sided auto-PSD is given as follows
\begin{equation}\label{Eq:ISO1}
{S_{d}}\left( \nu  \right) = {S_0}{\left( {\frac{\nu }{{{\nu _0}}}} \right)^{ - e}} \quad \mbox{for} \quad {\nu _a} \le \nu  \le {\nu_b},
\end{equation}
where $\nu_0 = 1$ rad/m denotes the reference angular spatial frequency and $S_0$ (m\textsuperscript{3}) is the amplitude of the PSD for $\nu=\nu_0$. A constant velocity auto-PSD can be obtained by imposing the undulation exponent $e$ equal to 2. When this occurrence is satisfied, the auto-PSD of surface-induced displacements is calculated through a simple integration of a flat-spectrum velocity signal. Rearranging Eq.~\eqref{Eq:ISO1}, the expression of the PSD in the angular frequency domain becomes  
\begin{equation}\label{Eq:ISO2}
{S_{d}}\left( \omega  \right) = \frac{{{S_d}\left( {\nu  = \omega /V} \right)}}{V} = \frac{{{S_0}}}{{{V}}}{\left( {\frac{\omega }{{{\omega _0}}}} \right)^{ - 2}}  \quad \mbox{for} \quad {\omega _a} \le \omega  \le {\omega _b},
\end{equation}
where $\omega _0=\nu _0 V$ indicates the reference angular frequency. 

Since in the classification of road profiles the angular spatial frequency usually ranges from $\nu_a=0.063$ rad/m to $\nu_b=17.77$ rad/m \cite{ISO8608} (corresponding to wavelengths from 100 m to 0.35 m, respectively), by considering the usual range of speed values from 10 to 30 m/s, we have that the angular frequency varies in the interesting range from $\omega_a=1.89$ rad/s (0.30 Hz) to $\omega_b=177.7$ rad/s (28.3 Hz). Thus, to account for these restrictions, we introduce lower and upper limits in Eqs.~\eqref{Eq:ISO1} and~\eqref{Eq:ISO2}. 

\begin{figure}[htbp]
\centering
\sidesubfloat[a]{\includegraphics[width=8.5cm]{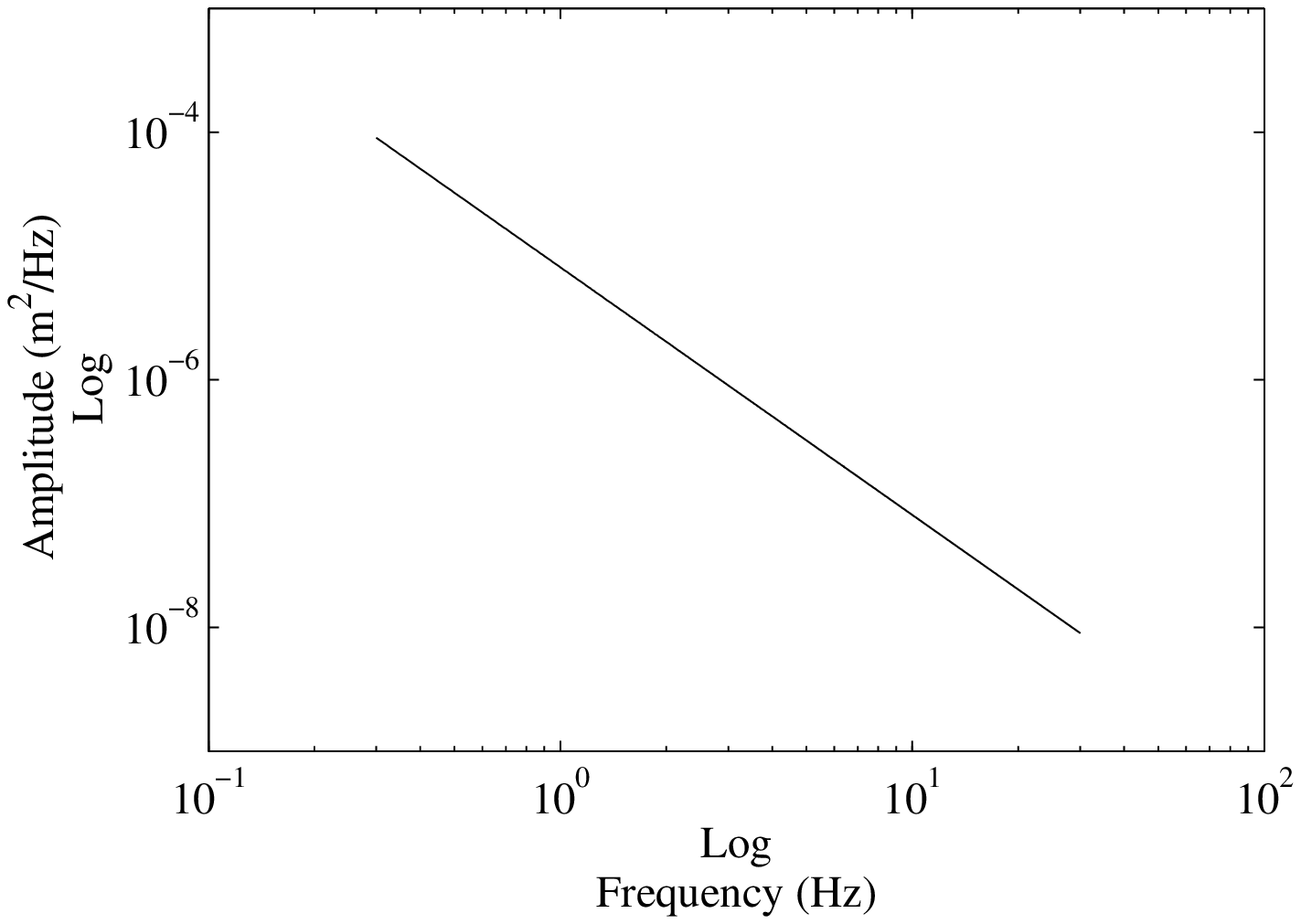}\label{Fig:Road_PSD}}
\sidesubfloat[b]{\includegraphics[width=8.5cm]{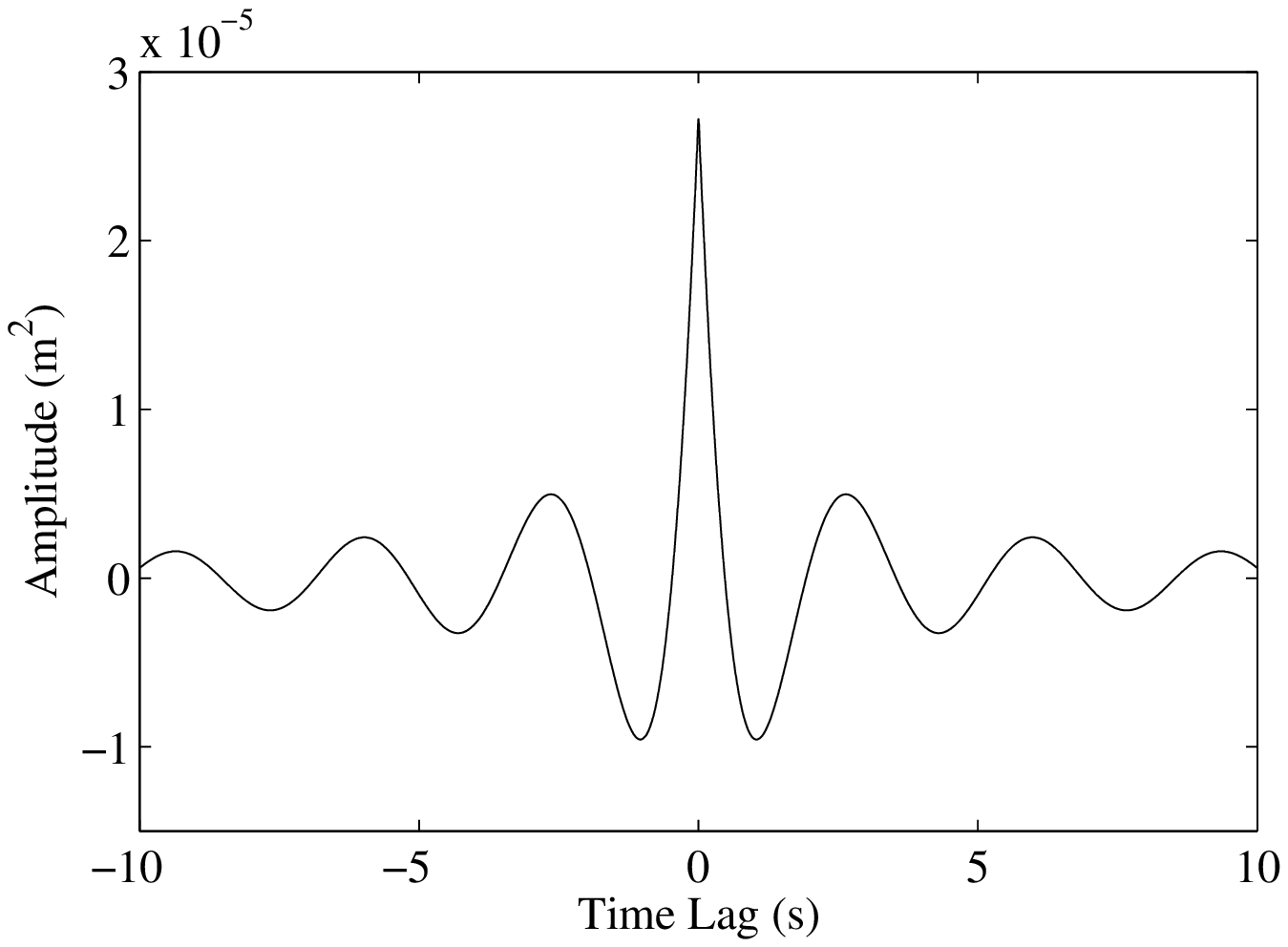}\label{Fig:Road_CorrFun}}
\caption{ISO 8608 - class C approximation ($\nu_0=1$ rad/m, $S_0=16\cdot10^{-6}$ m\textsuperscript{3}) for $V=20$ m/s: (a) auto-PSD and (b) auto-correlation function.}
\end{figure}

The ISO 8608 class C auto-PSD and auto-correlation function are shown in Figs.~\ref{Fig:Road_PSD} and~\ref{Fig:Road_CorrFun}, respectively. In general, an analytical or numerical expression of the auto-correlation function ${R_{d}}(\tau)$ can be obtained through inverse Fourier transformation of Eq.~\eqref{Eq:ISO2}. Specifically, the curve depicted in Fig.~\ref{Fig:Road_CorrFun}, representing ${R_{d}}(\tau)$ for the ISO 8608 case, is build up by numerically computing the IFT of ${S_{d}}(\omega)$.

With regards to the ordinary coherence function, we hypothesize that ${\it \Gamma_p}(\omega)$ in Eq.~\eqref{Eq:SiFPSDMat} have the trend of the fitting model proposed by Bogsj{\"o~\cite{Bogsjo2008}, and illustrated in Fig.~\ref{Fig:Road_CoheFun}. Experimental data acquired on very different road tracks have surprisingly shown a good agreement with this one-parameter exponentially decreasing function. At a constant velocity, the original exponential model can be rearranged in the angular frequency domain as
\begin{equation}\label{Eq:Cohe_Fun}
{\it \Gamma_p} \left( \omega  \right) = {{\rm{e}}^{\textstyle \frac{{ - \mu {W_p} \omega }}{{2\pi {V}}}}} \quad \mbox{for} \quad \omega \ge 0.
\end{equation} 
From practical considerations, it is basically possible to conclude that ${\it \Gamma_p}(\omega)$ has to approach 1 for long wavelengths (small $\omega$) and 0 for short wavelengths (large $\omega$). More over, as the two tracks become closer, that is $W_p \rightarrow 0$, then ${\it \Gamma_p}(\omega) \rightarrow 1$. The curve in Fig.~\ref{Fig:Road_CoheFun_IFT}, related to $H_p(\tau)$, is even build up by numerically computing the IFT of ${\it \Gamma_p}(\omega)$.
\begin{figure}[htbp]
\centering
\sidesubfloat[a]{\includegraphics[width=8.5cm]{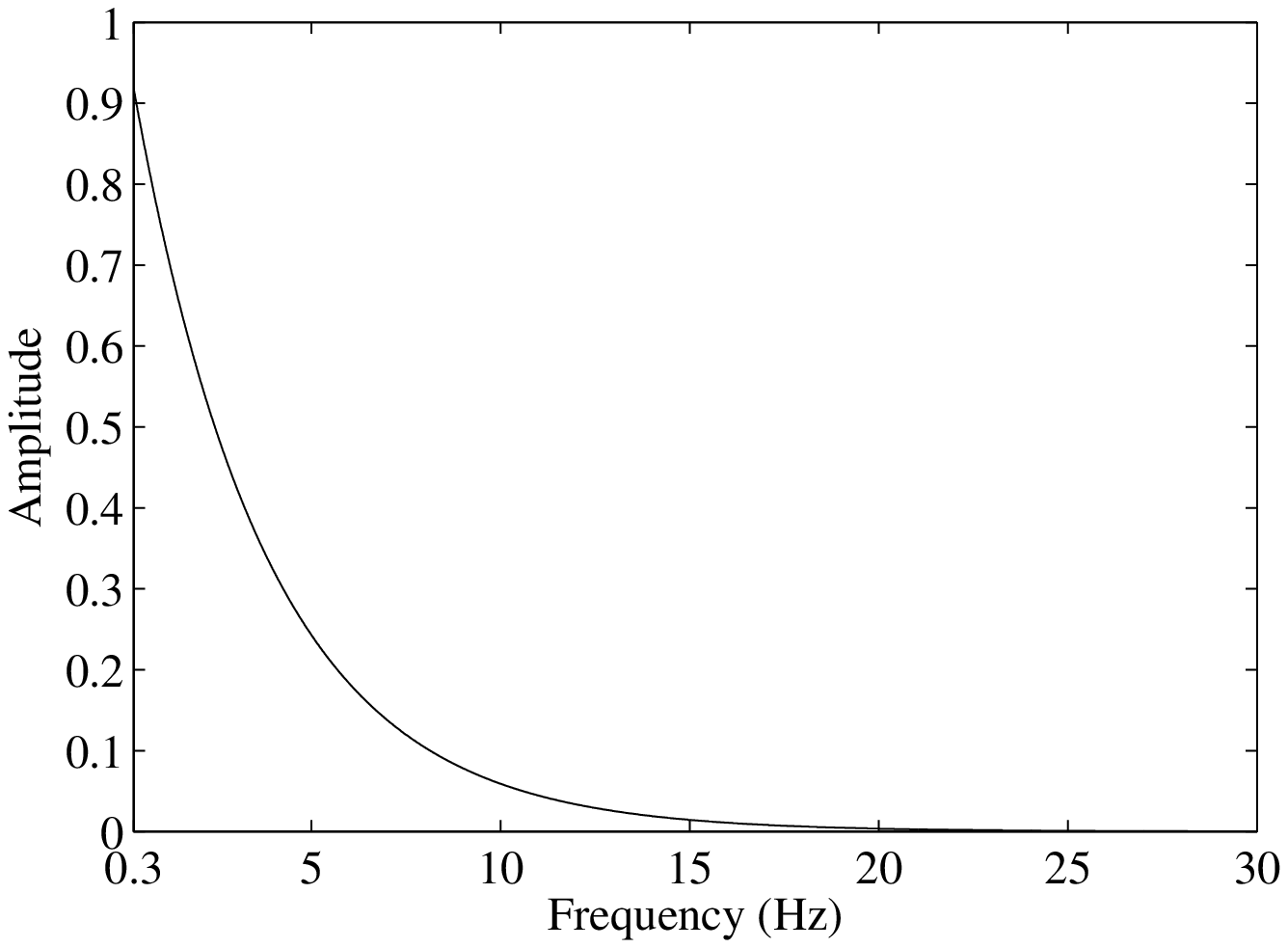}\label{Fig:Road_CoheFun}}
\sidesubfloat[b]{\includegraphics[width=8.5cm]{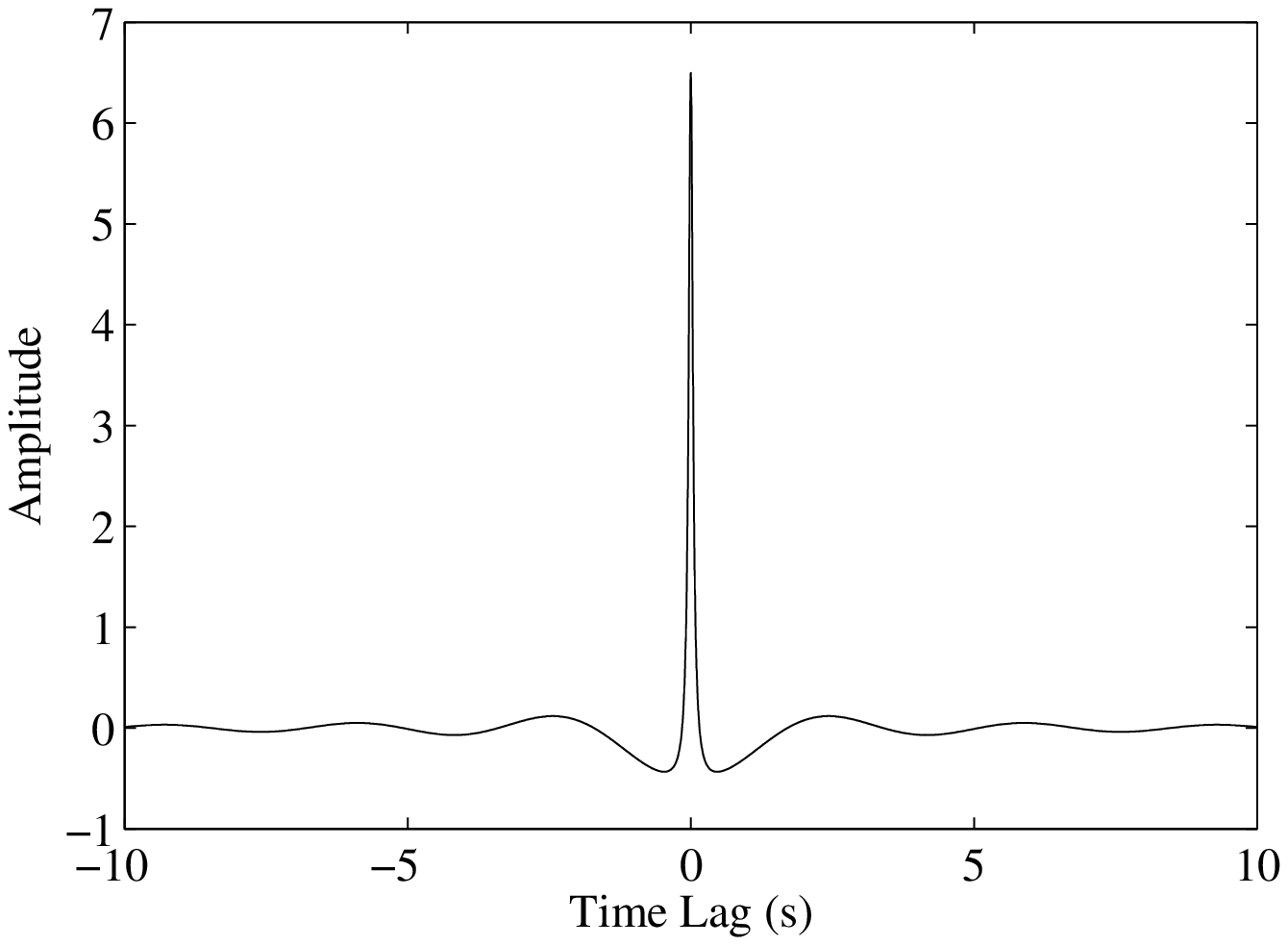}\label{Fig:Road_CoheFun_IFT}}
\caption{Bogsj{\"o} exponential model ($\mu=3.8$) for $V=20$ m/s: (a) ordinary coherence function and (b) its IFT.}
\end{figure}


\subsection{Analysis of the results}

By solving Eqs. from~\eqref{Eq:SSMatrices} to~\eqref{Eq:ModalVectors}, we compute resonance frequencies ($f_n$), damping ratios ($\zeta_n$) and modal vectors ($\boldsymbol{\psi}_n$) of the full-car model. The obtained modes of vibration are reported in Fig.~\ref{Fig:full-car_modes} for increasing values of the resonance frequency. While the computed modal parameters depend on the physical parameters of the exploited vehicle model, some considerations of general validity can be made, owing to well-known vehicle design procedures and to the fact that suspensions and wheels mounted on different axles have usually similar stiffness and damping values. Specifically, we notice the presence of high modal density and closely-spaced modes, that represent a first practical issue for the identification of vehicle systems. In the frequency band ranging from 0 to 20 Hz two main sets of rigid-body modes do exist. The three modes involving the chassis motion are all approximately located at 1 Hz, while the four modes related to rattle motions of front and rear axles are found in the range from 8 to 15 Hz depending on the vehicle parameter values (8 Hz in the case of values collected in Tab.~\ref{Tab:full-car_pars}). A second issue is caused by the presence of shock-absorbers with high amount of damping injected in the system, which is also a typical occurrence in vehicle systems.
\begin{figure}[htbp]
\subfloat[Heave: 1.064 Hz, 27.954 \%]{\includegraphics[width=5.5cm]{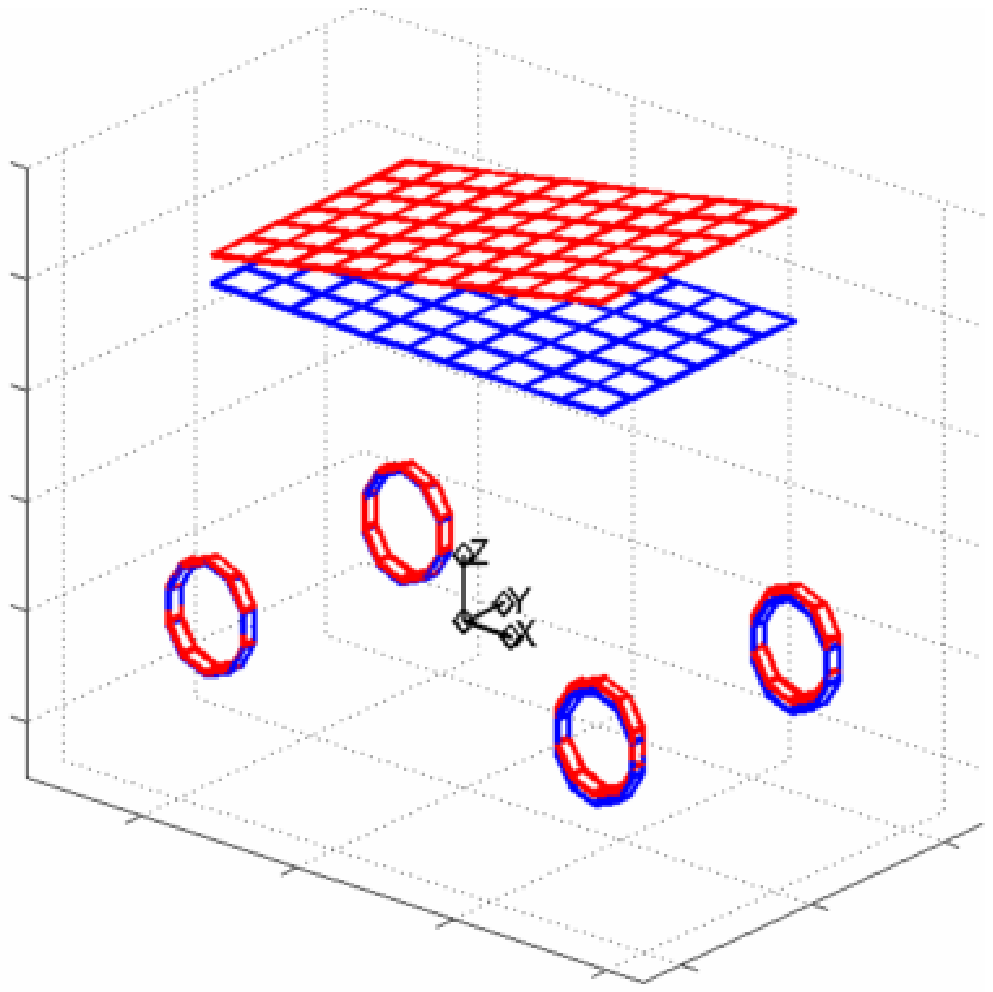}}
\hspace{0.5cm} 
\subfloat[Roll: 1.221 Hz, 32.189 \%]{\includegraphics[width=5.5cm]{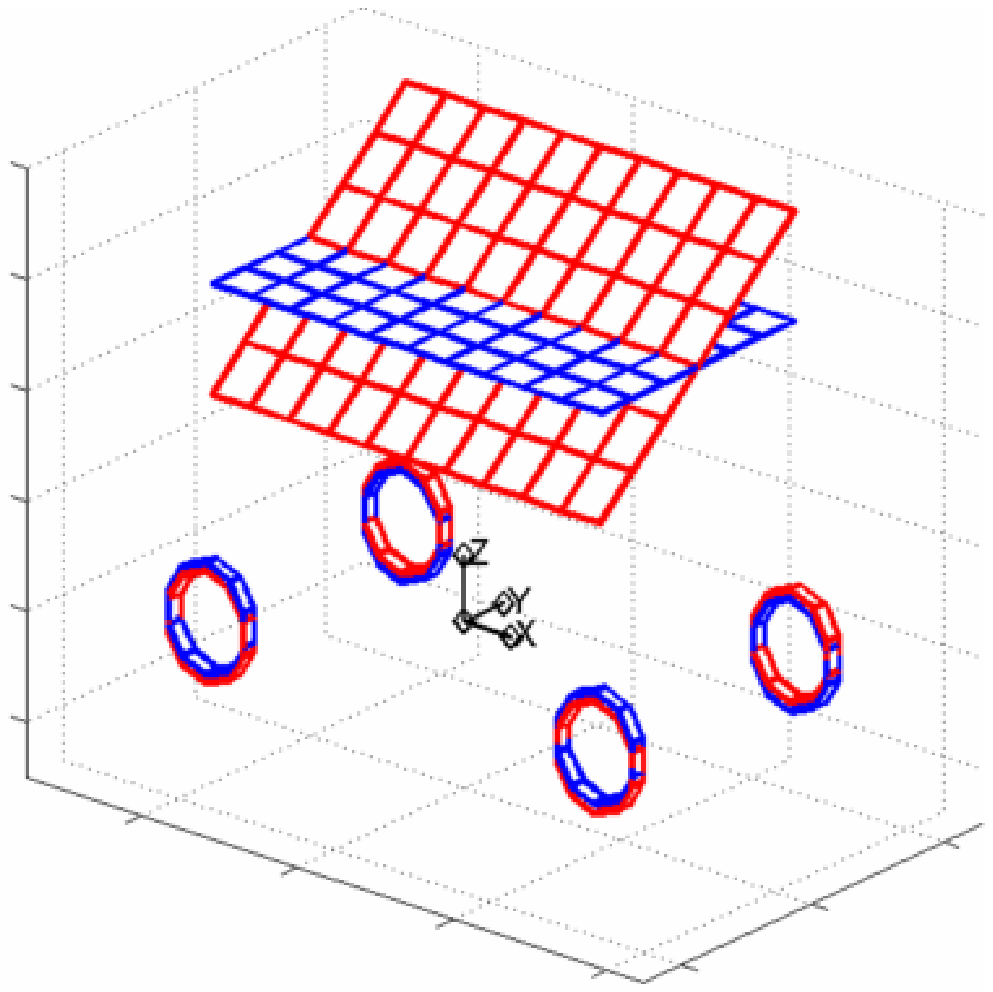}} 
\hspace{0.5cm} 
\subfloat[Pitch: 1.295 Hz, 34.295 \%]{\includegraphics[width=5.5cm]{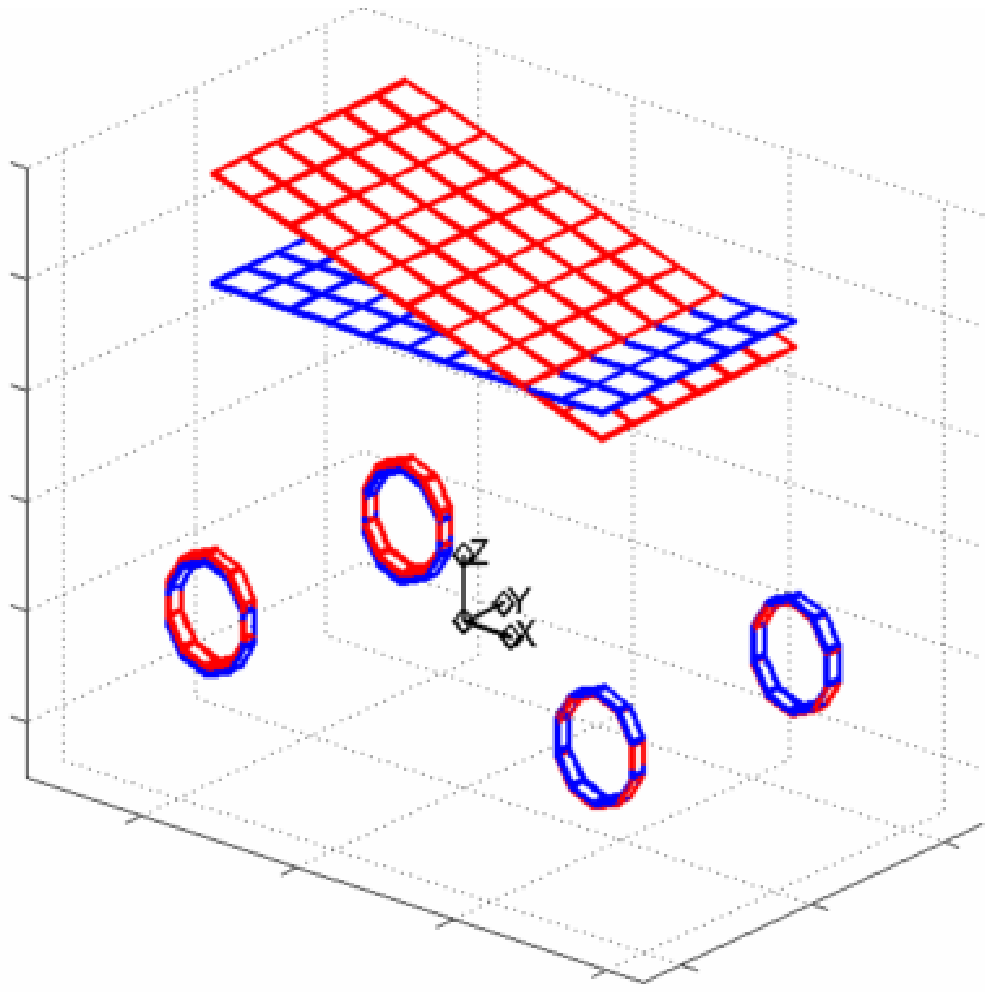}}\\
\vspace{0.5cm} 
\subfloat[Rear axle hop: 8.044 Hz, 26.031 \%]{\includegraphics[width=5.5cm]{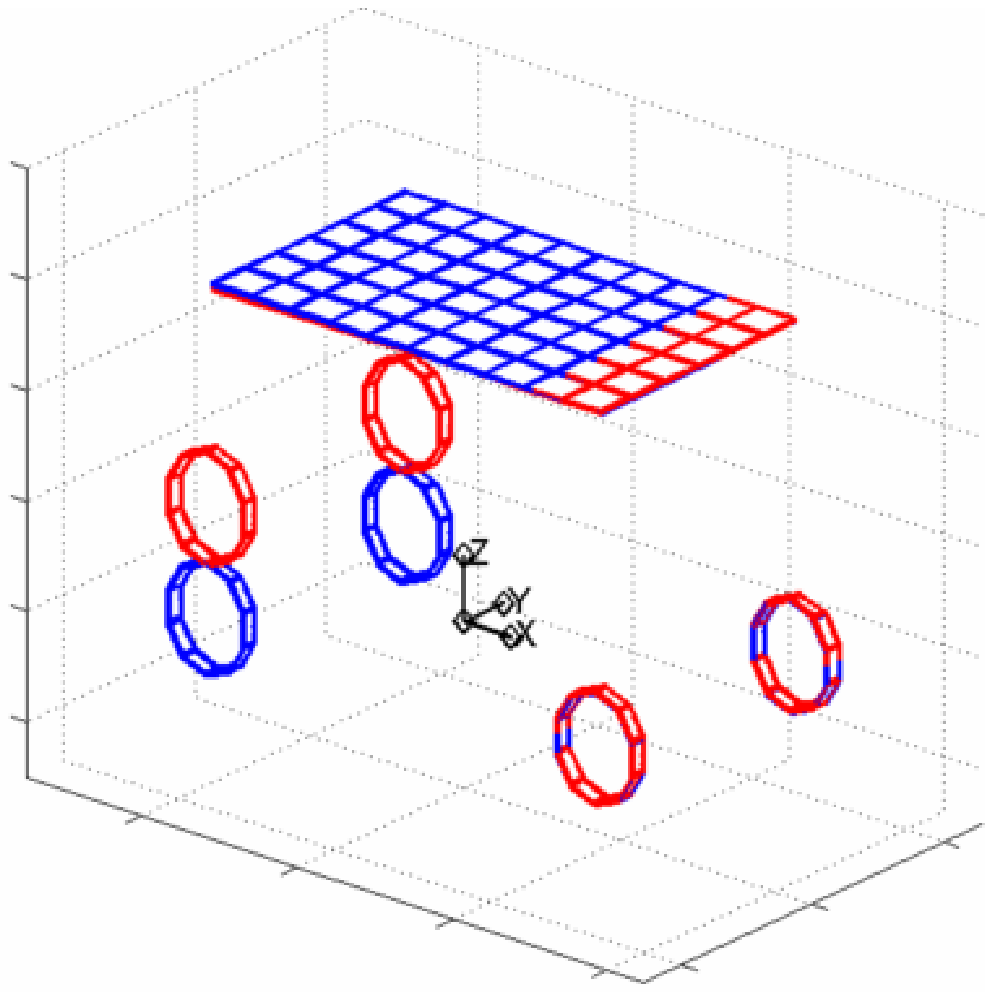}} 
\hspace{0.5cm} 
\subfloat[Front axle roll: 8.129 Hz, 27.794 \%]{\includegraphics[width=5.5cm]{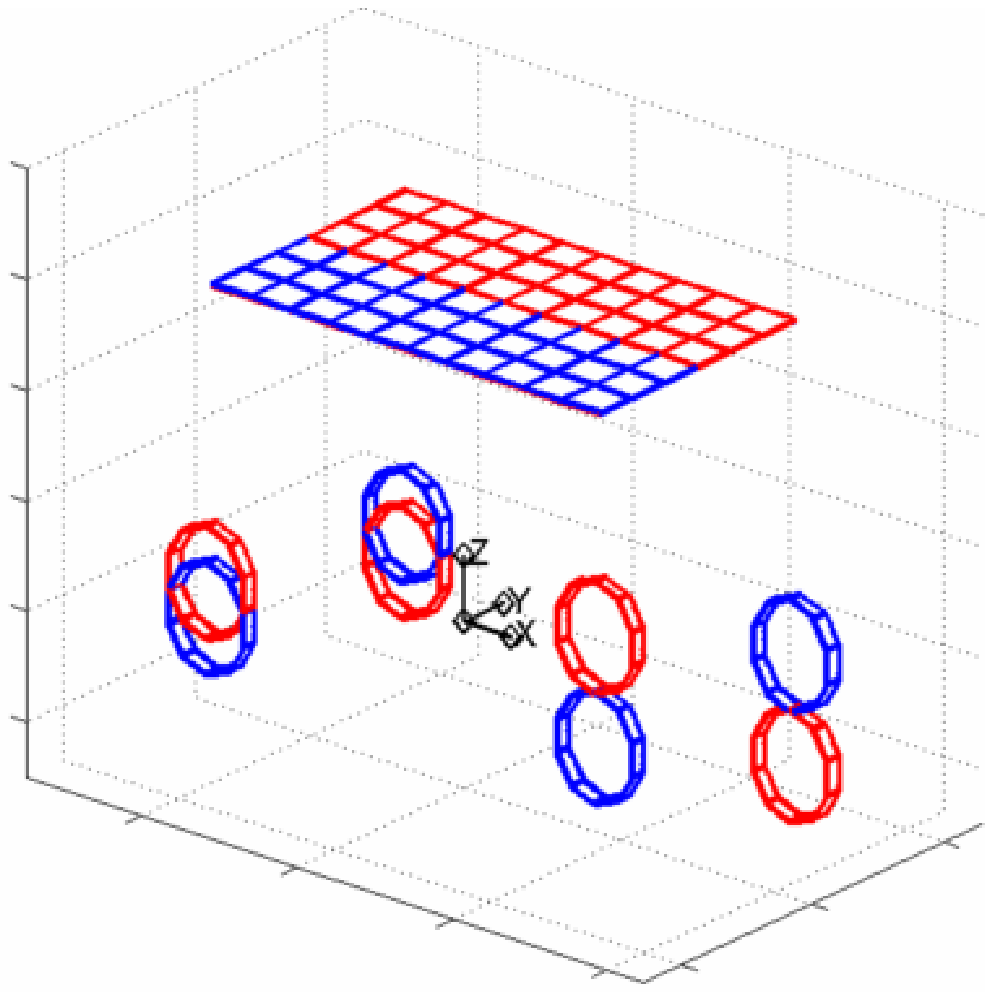}} 
\hspace{0.5cm} 
\subfloat[Front axle hop: 8.134 Hz, 28.372 \%]{\includegraphics[width=5.5cm]{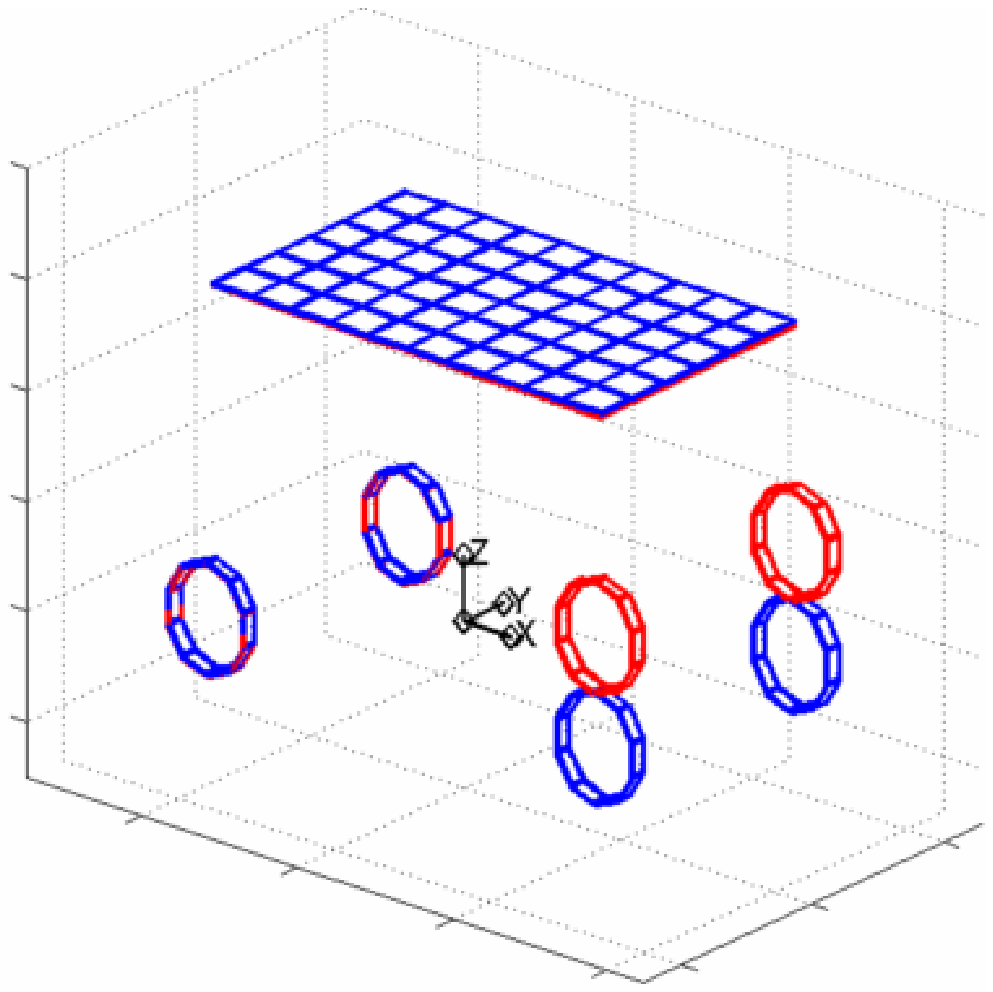}}\\
\vspace{0.5cm} 
\subfloat[Rear axle roll: 8.207 Hz, 25.354 \%]{\includegraphics[width=5.5cm]{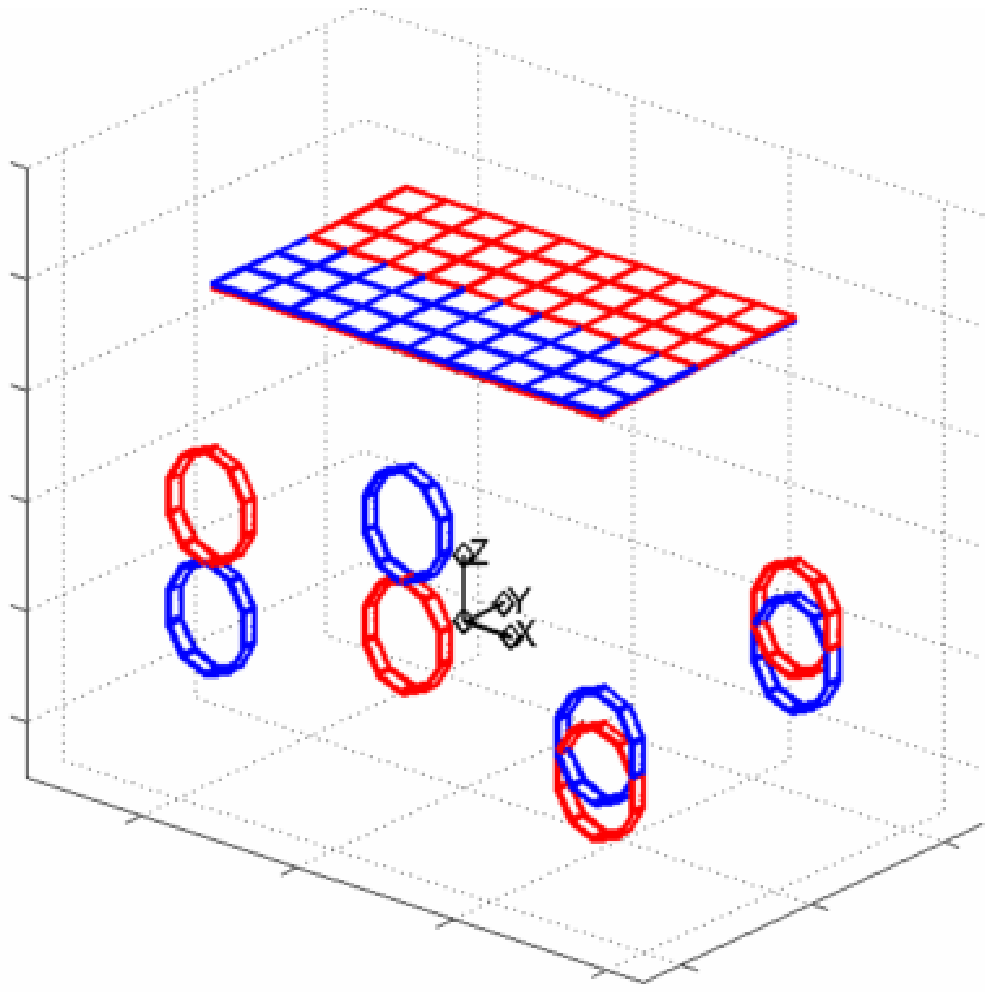}}
\caption{Modes of vibration of the full-car model. Blue and red lines indicate the undeformed and deformed shapes, respectively.}
\label{Fig:full-car_modes}
\end{figure}

In Fig.~\ref{Fig:Output_PSDs}, we show some output cross-PSDs of the full-car model. Owing to the high damping ratios, the resonance peaks, related to modes of vibration, are not clearly evidenced in the magnitude plots. More over, time and spatial correlations of surface-induced excitations produce relevant effects in form of distortions of cross-spectra. The time delay between the loads acting on the front and rear axles introduces the most significant of these distortions. In particular, we find (i) humps appearing in the magnitude of the cross-PSDs related to chassis dofs (Figs.~\ref{Fig:cog14psd} and~\ref{Fig:nocog23psd}) and (ii) the phase plots referred to the cross-PSDs of front and rear axles (Figs.~\ref{Fig:nocog46psd} and~\ref{Fig:nocog47psd}) revealing the typical saw-tooth trend caused by time-delayed signals. These distortions, well-known in the field of vehicle ride dynamics, are imputable to the so-called wheelbase filtering effect~\cite{Butkunas1966}. We comment that it should be expected that the presence in the output spectra of these distortion effects could hamper the correct operation of curve fitting methodologies based on the classical OMA modal model, that specifically relies on the NExT hypotheses.
\begin{figure}[htbp]
\centering
\sidesubfloat[a]{\includegraphics[width=8.5cm]{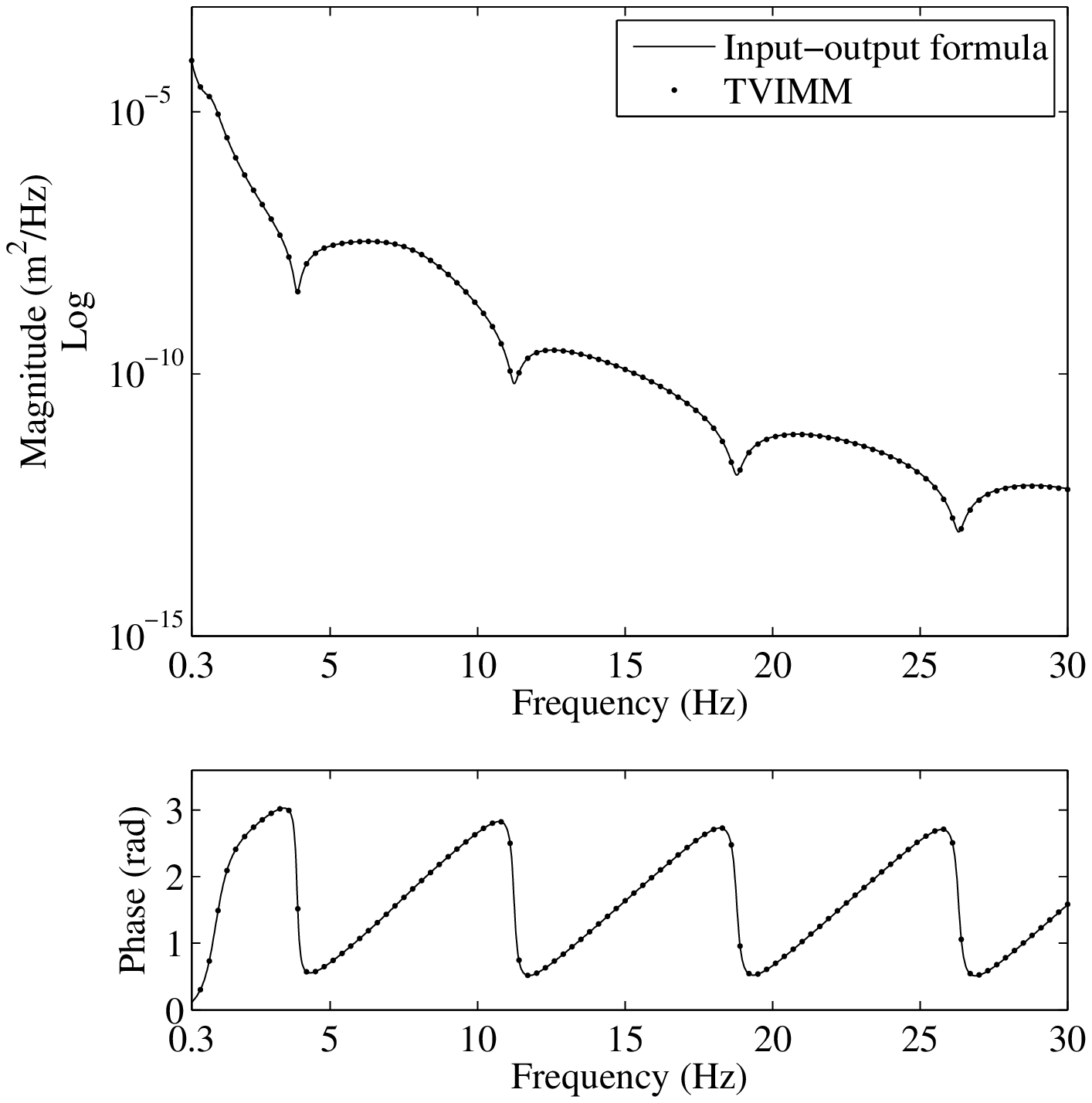}\label{Fig:cog14psd}}
\sidesubfloat[b]{\includegraphics[width=8.5cm]{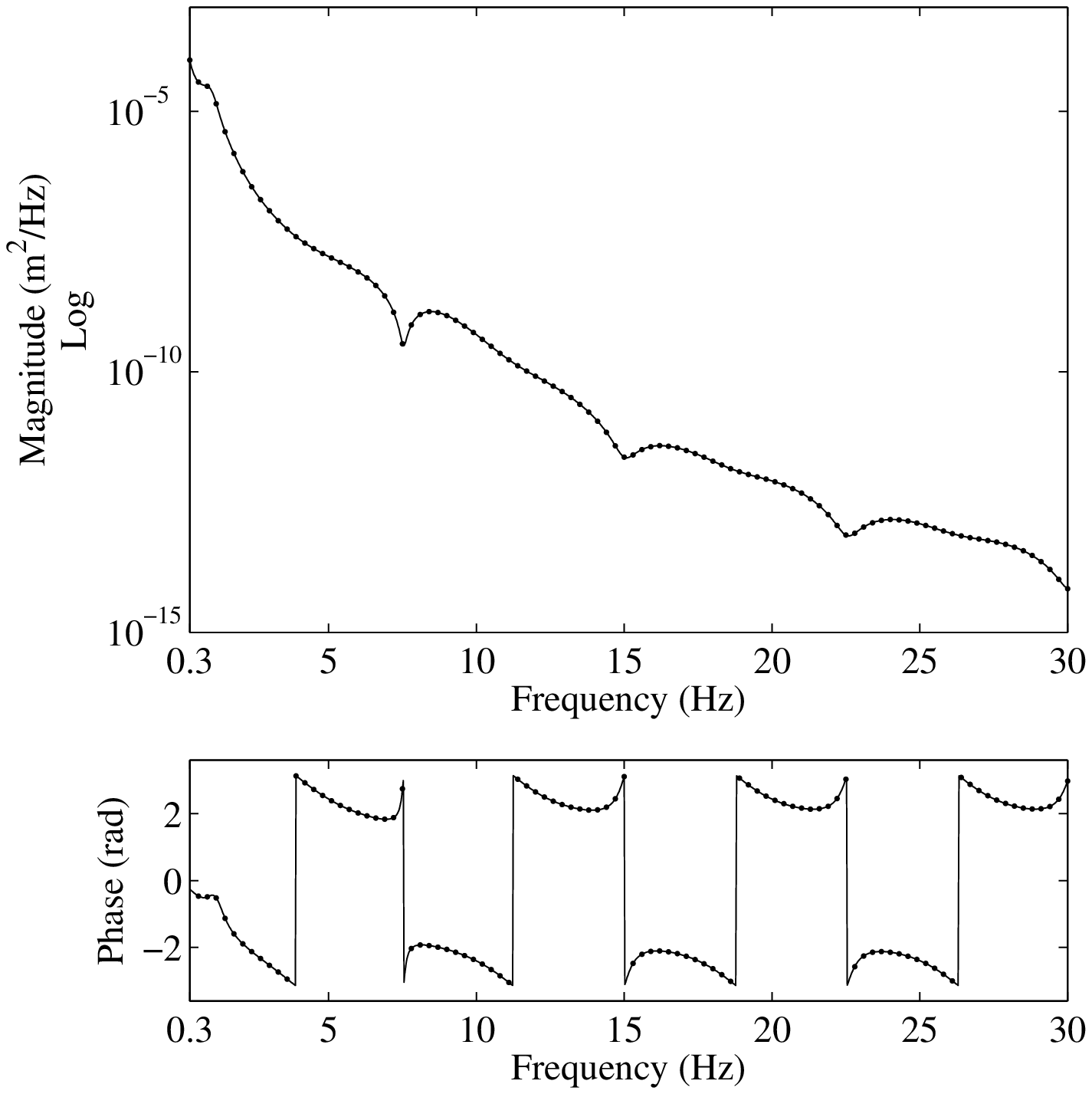}\label{Fig:nocog23psd}}\\
\vspace{0.5cm}
\sidesubfloat[c]{\includegraphics[width=8.5cm]{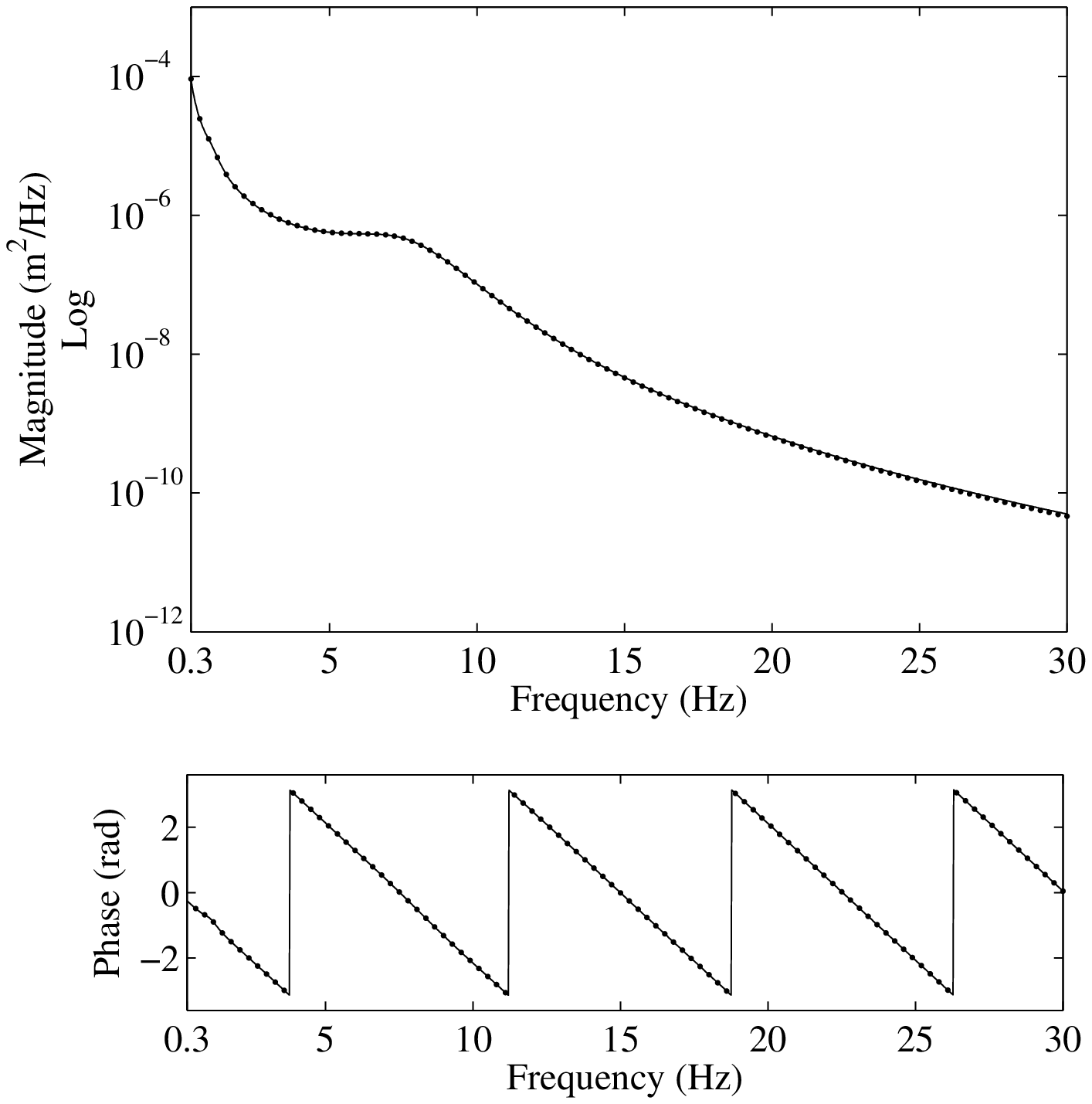}\label{Fig:nocog46psd}}
\sidesubfloat[d]{\includegraphics[width=8.5cm]{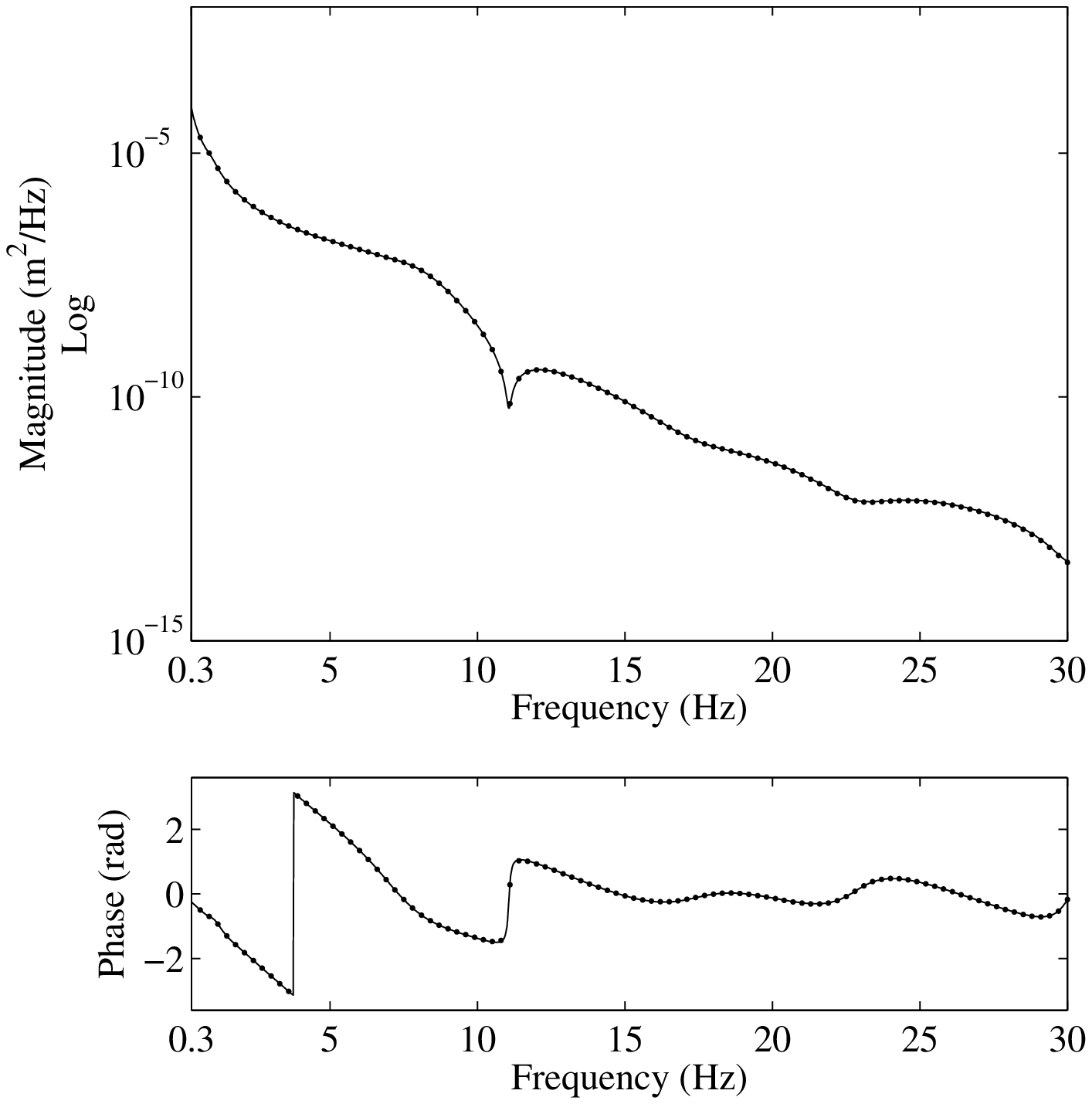}\label{Fig:nocog47psd}}\\
\caption{Output cross-PSDs of the full-car model (Fig.~\ref{Fig:Full-car}). Comparison between the input-output formula and the TVIMM: (a) $S_{z_{s}z_{u1}}(f)$, (b) $S_{z_{s2}z_{s3}}(f)$, (c) $S_{z_{u1}z_{u3}}(f)$ and (d) $S_{z_{u1}z_{u4}}(f)$.}
\label{Fig:Output_PSDs}
\end{figure}

By inspection of Fig.~\ref{Fig:Output_PSDs}, we find that the curves computed by using the proposed TVIMM perfectly overlap those obtained through the usage of the input-output formula. This proves the effectiveness of the proposed Track-Vehicle Interaction Modal Model in the frequency domain (Eq.~\eqref{Eq:PSDMat}). More over, some output cross-correlation functions of the full-car model are depicted in Fig.~\ref{Fig:Output_CorrFuns} and even in the case of time-lag domain (Eq.~\eqref{Eq:CorrMat}), we obtain the expected overlapping of the curves.
\begin{figure}[htbp]
\centering
\sidesubfloat[a]{\includegraphics[width=8.5cm]{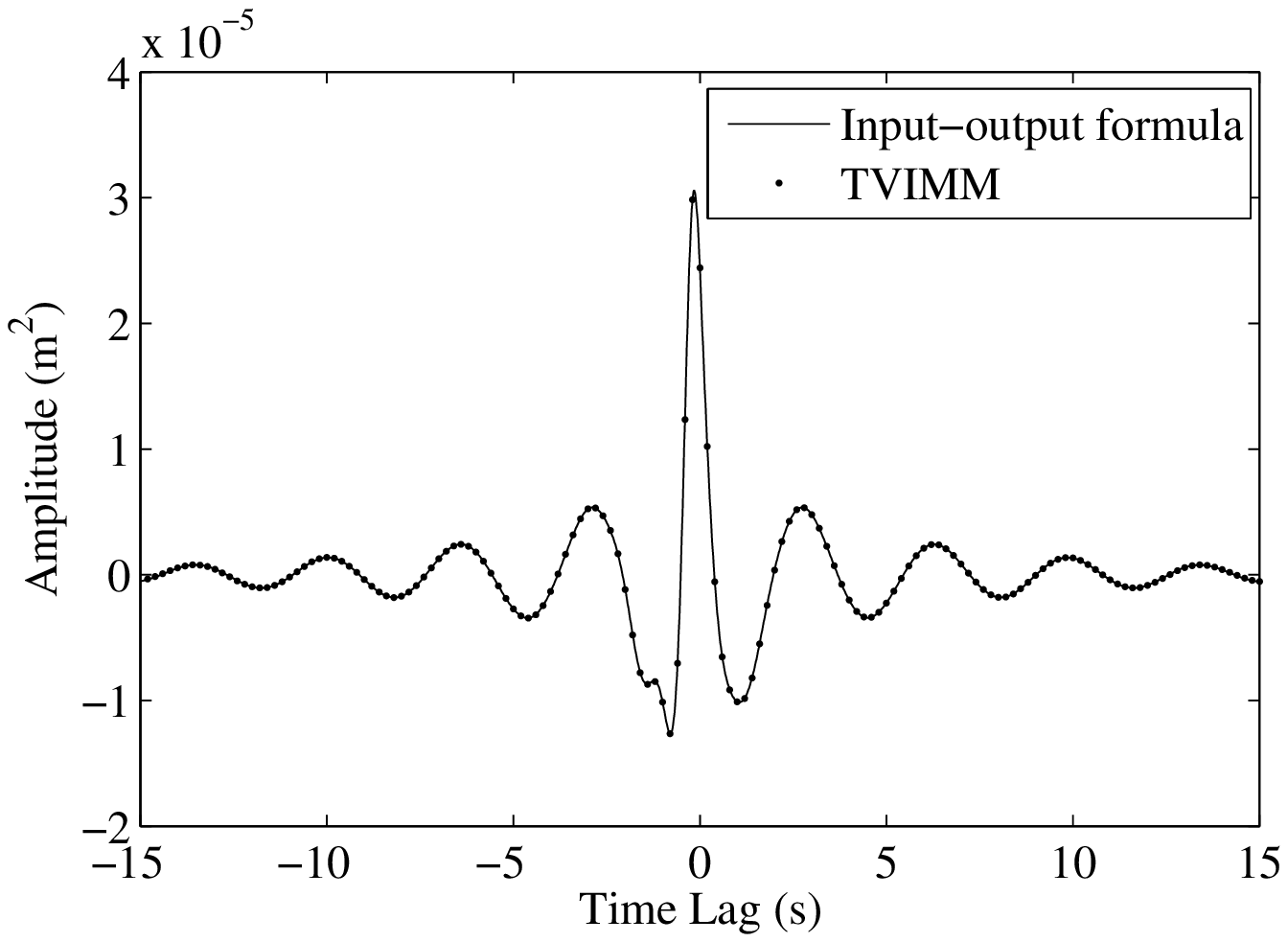}\label{Fig:cog14corr}}
\sidesubfloat[b]{\includegraphics[width=8.5cm]{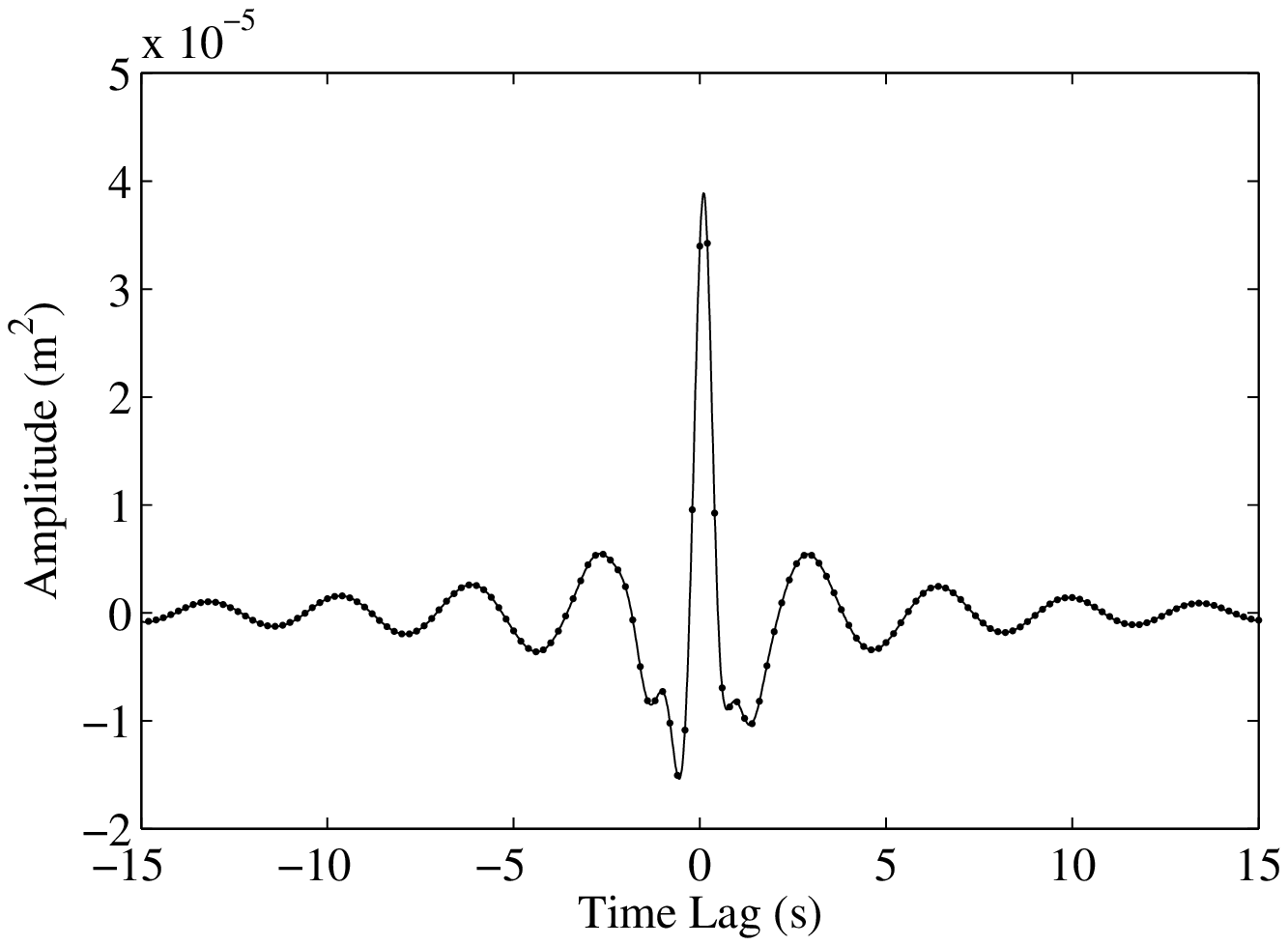}\label{Fig:nocog23corr}}\\
\vspace{0.5cm}
\sidesubfloat[c]{\includegraphics[width=8.5cm]{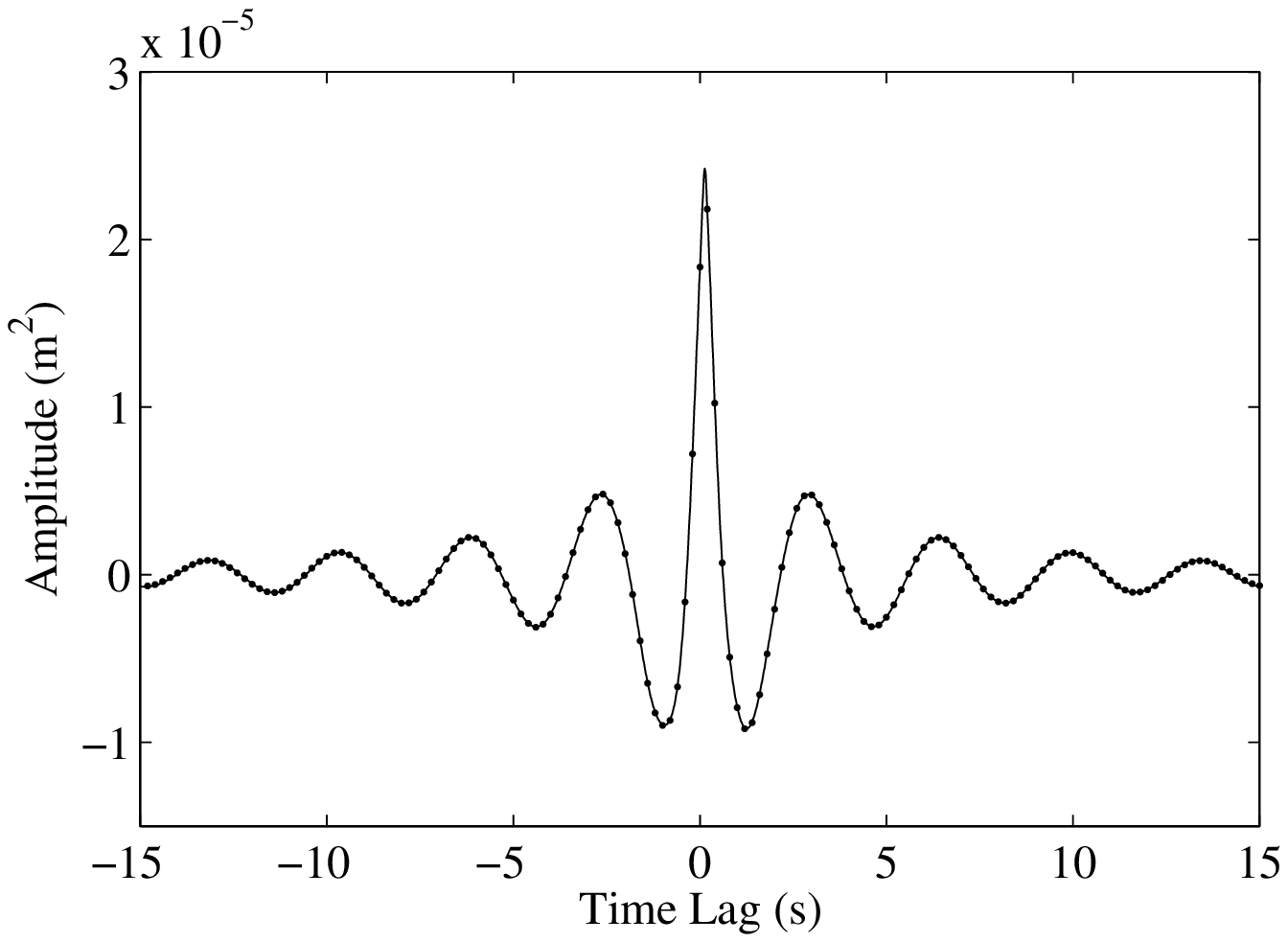}\label{Fig:nocog46corr}}
\sidesubfloat[d]{\includegraphics[width=8.5cm]{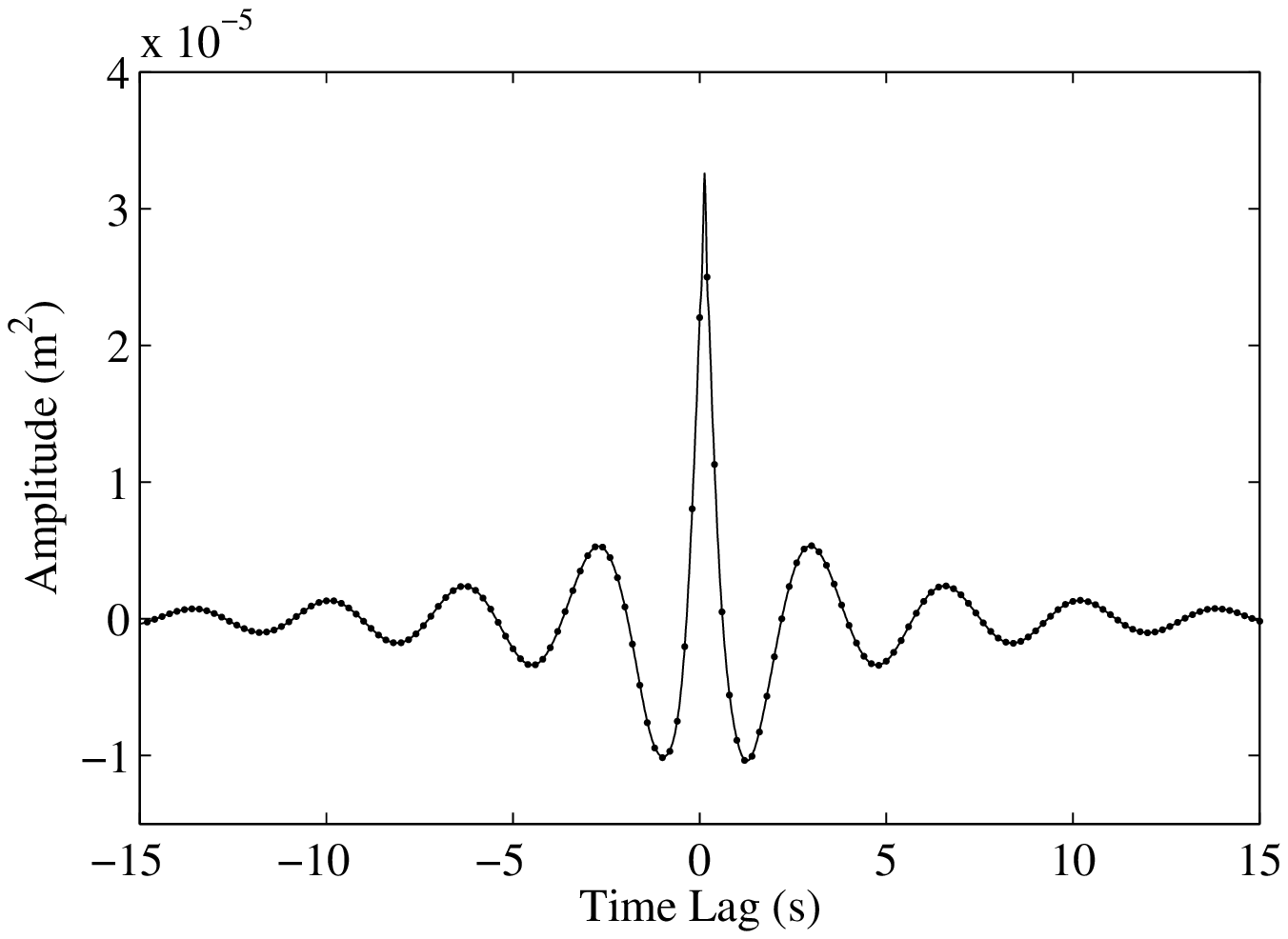}\label{Fig:cog14corr}}\\
\caption{Output cross-correlation functions of the full-car model (Fig.~\ref{Fig:Full-car}). Comparison between the input-output formula and the TVIMM: (a) $R_{z_{s}z_{u1}}(\tau)$, (b) $R_{z_{s2}z_{s3}}(\tau)$, (c) $R_{z_{u1}z_{u3}}(\tau)$ and (d) $R_{z_{u1}z_{u4}}(\tau)$.}
\label{Fig:Output_CorrFuns}
\end{figure}


\section{Conclusions}\label{section5}

Aim of the present paper is to introduce a novel OMA modal model, referred to as the Track-Vehicle Interaction Modal Model, needed for developing operational identification techniques, specifically designed for the case of vehicle systems and road/rail surface roughness. The effectiveness of the proposed formulation has been proved by simulating the response of a typical four-wheel vehicle model. Moving from the obtained expressions, specific estimation algorithms operating in the time-lag or in the frequency domain can be developed and implemented for practical purposes, providing new powerful tools to vehicle designers. By means of the resulting techniques, in fact, a map of vehicle system poles corresponding to the various operating conditions, as mainly at the different cruising speeds, can be achieved. The excitations produced during durability tests, in controlled conditions, could be utilised for providing the required operational loadings. Since these tests are usually already performed in scheduled test campaigns foreseen for vehicle design purposes, the usage of estimation techniques based on the modal model proposed in the paper would result simple and cost-effective. To this aim, both equipped laboratories and specific proving grounds could be successfully employed. As a further outcome of the proposed methodology, the resulting techniques could be used for the indirect characterisation of surface roughness, and, in turn, as a tool for improving safety and comfort.


\section*{Acknowledgments}

The authors Giovanni De Filippis and Davide Palmieri have attended a full Doctoral Program at Politecnico di Bari, with PhD scholarships funded, respectively, through the PON01\_02380 and the PON02\_00576\_3333585, by the Ministry of Education, University, and Research - MIUR, which is gratefully acknowledged.


\bibliography{mybibfile}

\begin{appendices}

\renewcommand{\theequation}{\Alph{section}\arabic{equation}}
\setcounter{equation}{0}

\section{Mathematical derivation of Eqs. from~\eqref{Eq:CorrMat} to~\eqref{Eq:PSDMat2}}\label{appendix}

Based on Eq.~\eqref{Eq:StaticGainMat}, we write the following equality
\begin{equation}
{\boldsymbol{\psi }}_n^T{{\boldsymbol{G}}_{fr}}\left( {{\boldsymbol{d}}\left( \tau  \right) \otimes {{\boldsymbol{\delta }}_p}\left( \tau  \right)} \right){\boldsymbol{G}}_{fr}^T{{\boldsymbol{\psi }}_m} = {\boldsymbol{\psi }}_n^{tT}\left( {\left( {{{\boldsymbol{K}}_t}{{\boldsymbol{\delta }}_p}\left( \tau  \right) - {{\boldsymbol{C}}_t}{{{\dot{\boldsymbol \delta }}}_p}\left( \tau  \right)} \right){{\boldsymbol{K}}_t} + \left( {{{\boldsymbol{K}}_t}{{{\dot {\boldsymbol \delta }}}_p}\left( \tau  \right) - {{\boldsymbol{C}}_t}{{{\ddot{\boldsymbol \delta }}}_p}\left( \tau  \right)} \right){{\boldsymbol{C}}_t}} \right){\boldsymbol{\psi }}_m^t,
\end{equation}
where ${\boldsymbol{\psi }}_n^t \in {\mathbb{C}^{1 \times N_t}}$ is a vector containing the modal components related to the wheels, that is, for the full-car model,
\begin{equation}
{\boldsymbol{\psi }}_n^t = {\left[ {\begin{array}{*{20}{c}}
{{\psi _{4n}}}&{{\psi _{5n}}}&{{\psi _{6n}}}&{{\psi _{7n}}}
\end{array}} \right]^T}.
\end{equation}
By substituting in Eq.~\eqref{Eq:R_qi_qj_2}, we get 
\begin{multline}
{R_{{q_i}{q_j}}}\left( \tau  \right) = {R_d}\left( \tau  \right)*\sum\limits_{n = 1}^{2N} {\sum\limits_{m = 1}^{2N} {\frac{{{\psi _{in}}{\psi _{jm}}}}{{{m_{an}}{m_{am}}}}} } \\
{\left( {\sum\limits_{p = 0}^{{N_w}} {{H_p}\left( \tau  \right)*} \left( {{\boldsymbol{\psi }}_n^{tT}\left( {\left( {{{\boldsymbol{K}}_t}{\boldsymbol{J}}_p^a\left( \tau  \right) - {{\boldsymbol{C}}_t}{\boldsymbol{J}}_p^b\left( \tau  \right)} \right){{\boldsymbol{K}}_t} + \left( {{{\boldsymbol{K}}_t}{\boldsymbol{J}}_p^b\left( \tau  \right) - {{\boldsymbol{C}}_t}{\boldsymbol{J}}_p^c\left( \tau  \right)} \right){{\boldsymbol{C}}_t}} \right){\boldsymbol{\psi }}_m^t} \right)} \right)},
\end{multline}
where the terms ${\boldsymbol{J}}_p^a(\tau)$, ${\boldsymbol{J}}_p^b(\tau)$ and ${\boldsymbol{J}}_p^c(\tau) \in {\mathbb{R}^{N_t \times N_t}}$ are matrices that encompass convolution integrals involving Dirac delta functions, located at different time-lag values (Eqs.~\eqref{Eq:DeltaMat1} and~\eqref{Eq:DeltaMat2}), and their derivatives. We rephrase the convolution integrals referred to ${\boldsymbol{J}}_p^a(\tau)$ as   
\begin{align}\label{Eq:Jpa}
\sum\limits_{p = 0}^{{N_w}} {{H_p}\left( \tau  \right)*} {\boldsymbol{J}}_p^a\left( \tau  \right) & = \sum\limits_{p = 0}^{{N_w}} {{H_p}\left( \tau  \right)*} \int\limits_{ - \infty }^{t + T} {\int\limits_{ - \infty }^t {{{\boldsymbol{\delta }}_p}\left( \tau  \right){{\rm{e}}^{{\lambda _n}\left( {t - \rho } \right)}}{{\rm{e}}^{{\lambda _m}\left( {t + \tau  - \sigma } \right)}}{\rm{d}}\rho {\rm{d}}\sigma}} \nonumber \\ 
 & = {H_0}\left( \tau  \right)*\left[ {\begin{array}{*{20}{c}}
{{E_1}\left( \tau  \right)}&0&{{E_3}\left( \tau  \right)}&0\\
0&{{E_1}\left( \tau  \right)}&0&{{E_3}\left( \tau  \right)}\\
{{E_2}\left( \tau  \right)}&0&{{E_1}\left( \tau  \right)}&0\\
0&{{E_2}\left( \tau  \right)}&0&{{E_1}\left( \tau  \right)}
\end{array}} \right] + {H_1}\left( \tau  \right)*\left[ {\begin{array}{*{20}{c}}
0&{{E_1}\left( \tau  \right)}&0&{{E_3}\left( \tau  \right)}\\
{{E_1}\left( \tau  \right)}&0&{{E_3}\left( \tau  \right)}&0\\
0&{{E_2}\left( \tau  \right)}&0&{{E_1}\left( \tau  \right)}\\
{{E_2}\left( \tau  \right)}&0&{{E_1}\left( \tau  \right)}&0
\end{array}} \right],
\end{align}
where the functions $E_1(\tau)$, $E_2(\tau)$ and $E_3(\tau)$ represent the explicit solutions of the convolution integrals, that is
\begin{equation}
{E_1}\left( \tau  \right) =  - \frac{{{{\rm{e}}^{{\lambda _m}\tau }}h\left( \tau  \right) + {{\rm{e}}^{ - {\lambda _n}\tau }}h\left( \tau  \right)}}{{{\lambda _n} + {\lambda _m}}},
\end{equation}
\begin{equation}\label{Eq:ExSol2}
{E_2}\left( \tau  \right) =  - \frac{{{{\rm{e}}^{{\lambda _m}\left( {\tau  + {\tau _1}} \right)}}h\left( {\tau  + {\tau _1}} \right) + {{\rm{e}}^{ - {\lambda _n}\left( {\tau  + {\tau _1}} \right)}}h\left( { - \tau  - {\tau _1}} \right)}}{{{\lambda _n} + {\lambda _m}}},
\end{equation}
\begin{equation}
{E_3}\left( \tau  \right) =  - \frac{{{{\rm{e}}^{{\lambda _m}\left( {\tau  - {\tau _1}} \right)}}h\left( {\tau  - {\tau _1}} \right) + {{\rm{e}}^{ - {\lambda _n}\left( {\tau  - {\tau _1}} \right)}}h\left( { - \tau  + {\tau _1}} \right)}}{{{\lambda _n} + {\lambda _m}}}.
\end{equation}
For example, Eq.~\eqref{Eq:ExSol2} can be proved by considering the following resolution scheme
\begin{align}\label{Eq:ExSol22}
{E_2}\left( \tau  \right) & = \int\limits_{ - \infty }^{t + \tau } {\int\limits_{ - \infty }^t {\delta \left( {\tau  + {\tau _1}} \right){{\rm{e}}^{{\lambda _n}\left( {t - \rho } \right)}}{{\rm{e}}^{{\lambda _m}\left( {t + \tau  - \sigma } \right)}}{\rm{d}}\rho {\rm{d}}\sigma } } \nonumber \\
& = \int\limits_{ - \infty }^{t + \tau } {{{\rm{e}}^{{\lambda _m}\left( {t + \tau  - \sigma } \right)}}{\rm{d}}\sigma \int\limits_{ - \infty }^t {\delta \left( {\sigma  - \rho  + {\tau _1}} \right){{\rm{e}}^{{\lambda _n}\left( {t - \rho } \right)}}{\rm{d}}\rho } } \nonumber \\
& = \int\limits_{ - \infty }^{t + \tau } {{{\rm{e}}^{{\lambda _m}\left( {t + \tau  - \sigma } \right)}}{{\rm{e}}^{{\lambda _n}\left( {t - \sigma  + {\tau _1}} \right)}}h\left( {t - \sigma  - {\tau _1}} \right){\rm{d}}\sigma }.
\end{align} 
Thus, by introducing the change of variable 
\begin{equation}
\bar \sigma  = \sigma  + {\tau _1} - t
\end{equation}
the last integral in Eq.~\eqref{Eq:ExSol22} can be solved as
\begin{align}
{E_2}\left( \tau  \right) & = {{\rm{e}}^{{\lambda _m}\left( {\tau  + {\tau _1}} \right)}}\int\limits_{ - \infty }^{\tau  + {\tau _1}} {{{\rm{e}}^{ - \left( {{\lambda _n} + {\lambda _m}} \right)\bar \sigma }}h\left( { - \bar \sigma } \right){\rm{d}}\bar \sigma } \nonumber \\
& = \int\limits_{ - \infty }^{\tau  + {\tau _1}} {{{\rm{e}}^{ - \left( {{\lambda _n} + {\lambda _m}} \right)\bar \sigma }}h\left( { - \bar \sigma } \right){\rm{d}}\bar \sigma }  = \left[ {\frac{{{{\rm{e}}^{ - \left( {{\lambda _n} + {\lambda _m}} \right)\bar \sigma }}}}{{ - \left( {{\lambda _n} + {\lambda _m}} \right)}}h\left( { - \bar \sigma } \right)} \right]_{ - \infty }^{\tau  + {\tau _1}} - \int\limits_{ - \infty }^{\tau  + {\tau _1}} {\frac{{{{\rm{e}}^{ - \left( {{\lambda _n} + {\lambda _m}} \right)\bar \sigma }}}}{{ - \left( {{\lambda _n} + {\lambda _m}} \right)}}\delta \left( { - \bar \sigma } \right){\rm{d}}\bar \sigma } \nonumber \\
& = \frac{{{{\rm{e}}^{ - \left( {{\lambda _n} + {\lambda _m}} \right)\left( {\tau  + {\tau _1}} \right)}}h\left( { - \tau  - {\tau _1}} \right)}}{{ - \left( {{\lambda _n} + {\lambda _m}} \right)}} + \int\limits_{ - \infty }^{\tau  + {\tau _1}} {\frac{{{{\rm{e}}^{ - \left( {{\lambda _n} + {\lambda _m}} \right)\bar \sigma }}}}{{ - \left( {{\lambda _n} + {\lambda _m}} \right)}}\delta \left( { - \bar \sigma } \right)\left( { - {\rm{d}}\bar \sigma } \right)} \nonumber \\
& = \frac{{{{\rm{e}}^{ - \left( {{\lambda _n} + {\lambda _m}} \right)\left( {\tau  + {\tau _1}} \right)}}h\left( { - \tau  - {\tau _1}} \right) + h\left( {\tau  + {\tau _1}} \right)}}{{ - \left( {{\lambda _n} + {\lambda _m}} \right)}}.
\end{align} 
Moving from Eq.~\eqref{Eq:Jpa}, we derive the expressions of the convolution integrals referred to ${\boldsymbol{J}}_p^b (\tau)$ and ${\boldsymbol{J}}_p^c (\tau)$, that are
\begin{equation}\label{Eq:Jpb}
\sum\limits_{p = 0}^{{N_w}} {{H_p}\left( \tau  \right)*} {\boldsymbol{J}}_p^b\left( \tau  \right) = \sum\limits_{p = 0}^{{N_w}} {{H_p}\left( \tau  \right)*} \int\limits_{ - \infty }^{t + T} {\int\limits_{ - \infty }^t {{{{\dot{\boldsymbol \delta }}}_p}\left( \tau  \right){{\rm{e}}^{{\lambda _n}\left( {t - \rho } \right)}}{{\rm{e}}^{{\lambda _m}\left( {t + \tau  - \sigma } \right)}}{\rm{d}}\rho {\rm{d}}\sigma } }  =  - {\lambda _n}\sum\limits_{p = 0}^{{N_w}} {{H_p}\left( \tau  \right)*} {\boldsymbol{J}}_p^a\left( \tau  \right),
\end{equation}
\begin{equation}\label{Eq:Jpc}
\sum\limits_{p = 0}^{{N_w}} {{H_p}\left( \tau  \right)*} {\boldsymbol{J}}_p^c\left( \tau  \right) = \sum\limits_{p = 0}^{{N_w}} {{H_p}\left( \tau  \right)*} \int\limits_{ - \infty }^{t + T} {\int\limits_{ - \infty }^t {{{{\ddot{\boldsymbol \delta }}}_p}\left( \tau  \right){{\rm{e}}^{{\lambda _n}\left( {t - \rho } \right)}}{{\rm{e}}^{{\lambda _m}\left( {t + \tau  - \sigma } \right)}}{\rm{d}}\rho {\rm{d}}\sigma }  = \lambda _n^2\sum\limits_{p = 0}^{{N_w}} {{H_p}\left( \tau  \right)*} {\boldsymbol{J}}_p^a\left( \tau  \right)}, 
\end{equation}
where the solutions of these convolution integrals can be retrieved from that of Eq.~\eqref{Eq:Jpa} by using the properties of Dirac delta function derivatives.

Eqs.~\eqref{Eq:Jpa},~\eqref{Eq:Jpb} and~\eqref{Eq:Jpc} lead to the following equality
\begin{align}
& \sum\limits_{p = 0}^{{N_w}} {{H_p}\left( \tau  \right)*} \left( {{\boldsymbol{\psi }}_n^{tT}\left( {\left( {{{\boldsymbol{K}}_t}{\boldsymbol{J}}_p^a\left( \tau  \right) - {{\boldsymbol{C}}_t}{\boldsymbol{J}}_p^b\left( \tau  \right)} \right){{\boldsymbol{K}}_t} + \left( {{{\boldsymbol{K}}_t}{\boldsymbol{J}}_p^b\left( \tau  \right) - {{\boldsymbol{C}}_t}{\boldsymbol{J}}_p^c\left( \tau  \right)} \right){{\boldsymbol{C}}_t}} \right){\boldsymbol{\psi }}_m^t} \right) = \nonumber \\
& = {H_0}\left( \tau  \right)*\left( {\left( {\left( {k_{ft}^2 - c_{ft}^2\lambda _n^2} \right)\left( {{\psi _{4m}}{\psi _{4n}} + {\psi _{5m}}{\psi _{5n}}} \right) + \left( {k_{rt}^2 - c_{rt}^2\lambda _n^2} \right)\left( {{\psi _{6m}}{\psi _{6n}} + {\psi _{7m}}{\psi _{7n}}} \right)} \right){E_1}\left( \tau  \right) + } \right. \nonumber \\
& + \left( {\left( {{k_{ft}} - {c_{ft}}{\lambda _n}} \right)\left( {{k_{rt}} + {c_{rt}}{\lambda _n}} \right)\left( {{\psi _{4m}}{\psi _{6n}} + {\psi _{5m}}{\psi _{7n}}} \right)} \right){E_2}\left( \tau  \right) + \nonumber \\
& = \left. { + \left( {\left( {{k_{ft}} + {c_{ft}}{\lambda _n}} \right)\left( {{k_{rt}} - {c_{rt}}{\lambda _n}} \right)\left( {{\psi _{6m}}{\psi _{4n}} + {\psi _{7m}}{\psi _{5n}}} \right)} \right){E_3}\left( \tau  \right)} \right) + \nonumber \\
& + {H_1}\left( \tau  \right)*\left( {\left( {\left( {k_{ft}^2 - c_{ft}^2\lambda _n^2} \right)\left( {{\psi _{4m}}{\psi _{5n}} + {\psi _{5m}}{\psi _{4n}}} \right) + \left( {k_{rt}^2 - c_{rt}^2\lambda _n^2} \right)\left( {{\psi _{6m}}{\psi _{7n}} + {\psi _{7m}}{\psi _{6n}}} \right)} \right){E_1}\left( \tau  \right) + } \right. \nonumber \\
& + \left( {\left( {{k_{ft}} - {c_{ft}}{\lambda _n}} \right)\left( {{k_{rt}} + {c_{rt}}{\lambda _n}} \right)\left( {{\psi _{4m}}{\psi _{7n}} + {\psi _{5m}}{\psi _{6n}}} \right)} \right){E_2}\left( \tau  \right) + \nonumber \\
& \left. { + \left( {\left( {{k_{ft}} + {c_{ft}}{\lambda _n}} \right)\left( {{k_{rt}} - {c_{rt}}{\lambda _n}} \right)\left( {{\psi _{6m}}{\psi _{5n}} + {\psi _{7m}}{\psi _{4n}}} \right)} \right){E_3}\left( \tau  \right)} \right).
\end{align} 
By substituting this expression in Eq.~\eqref{Eq:R_qi_qj_2}, we observe that the generic cross-correlation function $R_{q_{i}q_{j}}(\tau)$ can be rewritten as
\begin{align}\label{Eq:R_qi_qj_3}
{R_{{q_i}{q_j}}}\left( \tau  \right) & = {R_d}\left( \tau  \right)*\left( {\sum\limits_{p = 0}^{{N_w}} {{H_p}\left( \tau  \right)*} \left( {\sum\limits_{n = 1}^{2N} {\left( {\alpha _{in}^ph\left( \tau  \right) + \sum\limits_{l = 1}^{{N_L}} {\beta _{in}^{pl}{{\rm{e}}^{ + {\lambda _n}{\tau _l}}}h\left( {\tau  + {\tau _l}} \right) + \gamma _{in}^{pl}{{\rm{e}}^{ - {\lambda _n}{\tau _l}}}h\left( {\tau  - {\tau _l}} \right)} } \right){{\rm{e}}^{ + {\lambda _n}\tau }}{\psi _{jn}} + } } \right.} \right. \nonumber \\
& \left. {\left. { + \sum\limits_{n = 1}^{2N} {\left( {\bar \alpha _{jn}^ph\left( { - \tau } \right) + \sum\limits_{l = 1}^{{N_L}} {\bar \beta _{jn}^{pl}{{\rm{e}}^{ + {\lambda _n}{\tau _l}}}h\left( { - \tau  + {\tau _l}} \right) + \bar \gamma _{jn}^{pl}{{\rm{e}}^{ - {\lambda _n}{\tau _l}}}h\left( { - \tau  - {\tau _l}} \right)} } \right){{\rm{e}}^{ - {\lambda _n}\tau }}{\psi _{in}}} } \right)} \right),
\end{align} 
where the coefficients $\alpha _{in}^{pl}$, $\beta _{in}^{pl}$, $\gamma _{in}^{pl}$, $\bar{\alpha}_{in}^{pl}$, $\bar{\beta}_{in}^{pl}$, $\bar{\gamma}_{in}^{pl}$ play the same role as the operational reference factors $\phi_{in}$ in Eq.~\eqref{Eq:NExTCorrMat}. The number of coefficients in Eq.~\eqref{Eq:R_qi_qj_3} can be significantly decreased by recalling that
\begin{equation}
{R_{{q_i}{q_j}}}\left( \tau  \right) = {R_{{q_j}{q_i}}}\left( { - \tau } \right) \qquad \Leftrightarrow \qquad \alpha _{in}^{pl} = \bar \alpha _{in}^{pl} \qquad \beta _{in}^{pl} = \bar \beta _{in}^{pl} \qquad \gamma _{in}^{pl} = \bar \gamma _{in}^{pl}.
\end{equation}
Thus, for the full-car model we have
\begin{equation}
\alpha _{in}^0 = \sum\limits_{m = 1}^{2N} {\frac{{{\psi _{im}}\left( {\left( {c_{ft}^2\lambda _m^2 - k_{ft}^2} \right)\left( {{\psi _{4m}}{\psi _{4n}} + {\psi _{5m}}{\psi _{5n}}} \right) + \left( {c_{rt}^2\lambda _m^2 - k_{rt}^2} \right)\left( {{\psi _{6m}}{\psi _{6n}} + {\psi _{7m}}{\psi _{7n}}} \right)} \right)}}{{{m_{An}}{m_{Am}}\left( {{\lambda _n} + {\lambda _m}} \right)}}}, 
\end{equation}
\begin{equation}
\beta _{in}^{01} = \sum\limits_{m = 1}^{2N} {\frac{{{\psi _{im}}\left( {{c_{ft}}{\lambda _m} - {k_{ft}}} \right)\left( {{c_{rt}}{\lambda _m} + {k_{rt}}} \right)\left( {{\psi _{6m}}{\psi _{4n}} + {\psi _{7m}}{\psi _{5n}}} \right)}}{{{m_{An}}{m_{Am}}\left( {{\lambda _n} + {\lambda _m}} \right)}}}, 
\end{equation}
\begin{equation}
\gamma _{in}^{01} = \sum\limits_{m = 1}^{2N} {\frac{{{\psi _{im}}\left( {{c_{ft}}{\lambda _m} + {k_{ft}}} \right)\left( {{c_{rt}}{\lambda _m} - {k_{rt}}} \right)\left( {{\psi _{4m}}{\psi _{6n}} + {\psi _{5m}}{\psi _{7n}}} \right)}}{{{m_{An}}{m_{Am}}\left( {{\lambda _n} + {\lambda _m}} \right)}}}, 
\end{equation}
\begin{equation}
\alpha _{in}^1 = \sum\limits_{m = 1}^{2N} {\frac{{{\psi _{im}}\left( {\left( {c_{ft}^2\lambda _m^2 - k_{ft}^2} \right)\left( {{\psi _{4m}}{\psi _{5n}} + {\psi _{5m}}{\psi _{4n}}} \right) + \left( {c_{rt}^2\lambda _m^2 - k_{rt}^2} \right)\left( {{\psi _{6m}}{\psi _{7n}} + {\psi _{7m}}{\psi _{6n}}} \right)} \right)}}{{{m_{An}}{m_{Am}}\left( {{\lambda _n} + {\lambda _m}} \right)}}},
\end{equation}
\begin{equation}
\beta _{in}^{11} = \sum\limits_{m = 1}^{2N} {\frac{{{\psi _{im}}\left( {{c_{ft}}{\lambda _m} - {k_{ft}}} \right)\left( {{c_{rt}}{\lambda _m} + {k_{rt}}} \right)\left( {{\psi _{6m}}{\psi _{5n}} + {\psi _{7m}}{\psi _{4n}}} \right)}}{{{m_{An}}{m_{Am}}\left( {{\lambda _n} + {\lambda _m}} \right)}}}, 
\end{equation}
\begin{equation}
\gamma _{in}^{11} = \sum\limits_{m = 1}^{2N} {\frac{{{\psi _{im}}\left( {{c_{ft}}{\lambda _m} + {k_{ft}}} \right)\left( {{c_{rt}}{\lambda _m} - {k_{rt}}} \right)\left( {{\psi _{4m}}{\psi _{7n}} + {\psi _{5m}}{\psi _{6n}}} \right)}}{{{m_{An}}{m_{Am}}\left( {{\lambda _n} + {\lambda _m}} \right)}}}. 
\end{equation}
Finally, rearranging Eq.~\eqref{Eq:R_qi_qj_3} in a more suitable form we obtain
\begin{equation}
{R_{{q_i}{q_j}}}\left( \tau  \right) = \sum\limits_{n = 1}^{2N} {{{\bar \varphi }_{in}}\left( \tau  \right){\psi _{jn}}} {{\rm{e}}^{ + {\lambda _n}\tau }} + {\psi _{in}}{{\bar \varphi }_{jn}}\left( { - \tau } \right){{\rm{e}}^{ - {\lambda _n}\tau }},
\end{equation}
where ${\bar \varphi }_{in}\left( \tau  \right)$ is a lag-dependent operational reference factor, defined as
\begin{equation}
{{\bar \varphi }_{in}}\left( \tau  \right) = {R_d}\left( \tau  \right)*\sum\limits_{p = 0}^{{N_w}} {{H_p}\left( \tau  \right)*} \left( {\alpha _{in}^ph\left( \tau  \right) + \sum\limits_{l = 1}^{{N_L}} {\beta _{in}^{pl}{{\rm{e}}^{ + {\lambda _n}{\tau _l}}}h\left( {\tau  + {\tau _l}} \right) + \gamma _{in}^{pl}{{\rm{e}}^{ - {\lambda _n}{\tau _l}}}h\left( {\tau  - {\tau _l}} \right)} } \right).
\end{equation}

\end{appendices}

\end{document}